# What Do Chinese-Language Generative Search Engines Cite and Surface? A Large-Scale Empirical Study

Tao Zhen†, Yue Liu†*, Gege Zhang‡, Yixuan Niu‡

Aidso Wendao Research Institute, Beijing Aichacha Technology Co., Ltd., Beijing, China

† These authors contributed equally and share first authorship.
‡ These authors contributed equally and share second authorship.
* Corresponding author: badaoliu@adsida.com

# Contents

# Abstract

Generative AI question-answering systems have become an important gateway through which users access information, extending the study of content visibility from ranking in traditional search results to citations and presentation within generated answers. Generative engine optimization (GEO) refers to the study of factors associated with content retrieval, citation, and presentation in large-language-model-powered search systems. Emerging in response to this shift, GEO examines the conditions under which content is retrieved, cited, and presented in generative answers. Observable evidence on these processes in Chinese-language generative engines remains limited. Using a controlled data-collection design, this study covered the Web and App interfaces of four mainstream Chinese large-language-model platforms—eight platform interfaces in total—and 614 controlled queries; each query–platform-interface combination was collected in three fixed replications. The raw dataset comprised 214,119 records. After field linkage, deduplication, and removal of invalid records, a citation-level master dataset of 160,860 records was constructed and used to analyze the distributions and statistical associations of citations, entity exposure, and cross-interface consistency.

This study focuses on entity exposure during generation and treats the characteristics of cited sources and the degree of citation absorption as observational explanatory variables. Five principal findings emerged. First, brands in the citation pool were highly selectively included in answers, with an overall selection rate of 8.3%; 12.4% of the retrieved sources containing contact information contributed contact information to the answers. Second, content fit, cross-source occurrence count, and semantic role had relatively high importance in the corresponding predictive models. The 5118-Baidu Composite Quality Score was constructed from the 5118 and Baidu Webmaster site ratings, which are core reference metrics in traditional SEO. This composite variable was not the leading predictor for any of the outcomes examined; it was weakly associated with citation absorption and within-answer position and was a moderately positive predictor in the brand-selection model. Third, among cited pages with publication dates, the fitted half-lives for high- and low-timeliness queries were approximately 39 and 68 days, respectively, indicating that cited pages for highly time-sensitive queries were more concentrated in recent content. Fourth, approximately 13% of brand exposures could not be matched to the contemporaneous citation pool, while approximately 71% of contact-information exposures could not be matched to the crawled body text, suggesting that some entities presented in answers were not covered by the citation text visible at the time. Fifth, source sets differed systematically between the App and Web interfaces of the same platform, indicating that the access interface is an important dimension of stratification.

Keywords: generative search; information retrieval; citation behavior; source attribution; entity exposure; large language models; generative engine optimization (GEO)

# Introduction

## Research Background

Over the past two decades, search engine optimization (SEO) has primarily examined the ranking of webpages in search results and the factors that influence those rankings. With the adoption of generative question-answering products such as ChatGPT, DeepSeek, Doubao, and Qwen, users can also obtain information directly from natural-language answers. Research on visibility has therefore expanded beyond search ranking to whether a source is listed, whether its content is absorbed into an answer, and whether relevant entities appear in that answer. For brands and content providers, inclusion in a citation list and entity exposure in the body of an answer are two distinct stages. Research on optimizing citations and presentation in generated answers is commonly referred to as generative engine optimization (GEO).

## Problems and Challenges

Existing GEO research has examined content visibility, source citation, and optimization methods in generated answers, but systematic comparisons across multiple Chinese platforms using a unified field schema and citation-level data remain limited. Using platform-returned sources, answer text, and interface elements as observational material, this study addresses three questions: how cited sources and their degree of citation absorption are distributed; under what conditions brands and contact information enter answers; and whether the source sets returned through different interfaces of the same platform are consistent. The analyses distinguish citation-level, answer-level, and platform-interface-level units of analysis and report both descriptive distributions and predictive associations.

## Research Questions and Contributions

This study addresses the following three research questions:

1. Among cited webpages, what distributions and associations characterize source type, degree of citation absorption, semantic role, within-answer position, and article recency?

2. Within the controlled query set, how is the inclusion of brands and contact information in answers statistically associated with cross-source occurrence count, degree of citation absorption, semantic role, and query design, and how are exposures that cannot be matched to the contemporaneous citations or crawled body text distributed?

3. How consistent are the source sets and exposure outcomes between the App and Web interfaces of the same platform, and to which platforms, interfaces, and industry domains do the comparative results apply?

In addressing these research questions, this study makes three contributions:

1. It constructs a unified citation-level measurement corpus for Chinese-language generative engines. The study covers eight platform interfaces, 614 queries, and three fixed collection replications. The raw data comprise 214,119 records, which were cleaned to form a citation-level master dataset of 160,860 records. Under a unified field schema, the data-processing pipeline links answers, citations, and source attributes and includes a full crawl of the original text of cited sources; video and image materials were transcribed and incorporated into the textual analysis.

2. It proposes an analytical framework that distinguishes citation selection, citation absorption, entity exposure, and interface consistency. Nominal, general, and deep citations are used as operational categories for the extent to which source content is used in an answer, with distinct units of analysis specified at the citation, answer, and platform-interface levels. This produces a four-layer measurement framework with explicit units of analysis and verifiable indicators.

3. Under a unified measurement framework, it compares the statistical associations of the 5118-Baidu Composite Quality Score, article recency, cross-source occurrence count, semantic role, and platform interface with citation and exposure outcomes. The study reports effect direction, predictive importance, and robustness tests and, in accordance with the research design, separately estimates between-group differences in query formulations and conditional associations with source attributes.

# Related Work

## From SEO to GEO

Traditional information-retrieval and SEO research has developed relatively mature analytical frameworks around ranking signals such as content relevance, link authority, and click behavior (Manning et al., 2008). The rise of generative engines has extended the target of optimization from ranking in search results to citation and presentation in generated answers, giving rise to the emerging field of GEO. Aggarwal et al. (2024) were the first to systematically formulate generative engine optimization, defining it as "the first novel paradigm to aid content creators in improving their content visibility in generative engine responses" (Aggarwal et al., 2024, p. 5). Toward this goal, Lüttgenau et al. (2025) examined fine-grained tuning for domain-specific content and its performance in generative search results. Alpar et al. (2026) further argued that, beyond keyword generation, GEO must jointly consider content structure, textual distinctiveness, and technical implementation, while adapting editorial and technical strategies to the way generative systems evaluate content. Qi (2024) connected the design of pretrained generative models and improvements in inference efficiency with generative retrieval, discussing content relevance and presentation effects in search advertising. Collectively, these studies indicate that visibility in generative environments cannot be fully explained by conventional ranking metrics. In light of this theoretical shift, this study uses the 5118-Baidu Composite Quality Score as an operational proxy for traditional SEO performance while treating whether a source is cited, the degree to which its content is absorbed into an answer, and whether an entity enters the answer as separate outcome variables. It thereby evaluates the relative explanatory power of traditional SEO signals and variables such as content fit, article recency, and cross-source occurrence count on Chinese platforms. Existing GEO studies have largely focused on English-language platforms, practical guidance, or small samples; this study instead provides citation-level evidence from multiple Chinese platforms under a unified field schema.

The concurrent work most closely related to this study is the two-stage GEO measurement framework proposed by Zhang et al. (2026). It distinguishes "citation selection, where a platform triggers search and chooses sources, and citation absorption, where a cited page contributes language, evidence, structure, or factual support to the final answer" (Zhang et al., 2026, p. 1), primarily for English-language platforms such as ChatGPT, Gemini, and Perplexity. This distinction provides a direct theoretical basis for the analytical levels used here: a source entering the citation list constitutes citation selection, whereas the extent to which source content is actually used in the answer constitutes citation absorption. This study further operationalizes citation absorption into nominal, general, and deep categories and treats brand and contact-information exposure in the answer as a distinct outcome layer following citation. In contrast to Zhang et al. (2026), the present study focuses on Chinese-language generative engines, including Doubao, DeepSeek, Tencent Yuanbao, and Qwen. In addition to citation selection and absorption, it examines brand exposure, exposures unmatched to the contemporaneous citations, and contact-information exposure; compares the consistency of source sets between App and Web interfaces; and analyzes associations with article recency, cross-source occurrence count, and the 5118-Baidu Composite Quality Score. It therefore extends the two-stage framework with two additional measurement layers—entity exposure and interface consistency—and defines the unit of analysis for each layer.

With respect to brand visibility, Kumar (2026) observed a hierarchy by brand stature in English-language AI search engines: "global household names (e.g., Stripe, Nike) appear in 73% of relevant AI answers on their first run; established mid-market and regional brands (e.g., Olipop, Klaviyo) in 44%; niche and small brands in just 11%" (Kumar, 2026, p. 1). This finding suggests that preexisting brand recognition may be associated with the opportunity to appear in generated answers and that total brand frequency should be treated as an important control when analyzing differences in exposure. The Chinese-platform sample analyzed here likewise exhibited strong selectivity: the overall selection rate for brands in citation pools was 8.3%. Because the platforms, query sets, brand definitions, and denominators differ, however, this percentage is not compared directly in magnitude with the rate reported by Kumar (2026). Instead, the citation-level data distinguish total brand frequency, cross-source occurrence count, degree of citation absorption, and semantic similarity score. Alipour et al. (2026) found that generative-search citations “exhibit exposure bias toward already prominent voices, which risks entrenching incumbents and narrowing viewpoint diversity.” This provides a theoretical motivation for simultaneously controlling for brand frequency and testing cross-source occurrence count. Based on 11,500 real-world queries, Grossman et al. (2026) found that approximately 51.5% of representative queries triggered an AI Overview positioned above organic results, indicating that generative presentation has itself become a visibility gateway distinct from traditional ranking. Accordingly, the present study analyzes not only whether a source enters the citation list but also whether a brand actually enters the answer as a separate outcome. Other research, including Nimase et al.'s (2026) GEO-Bench, has examined adversarial manipulation of GEO and its detectability. Whereas that line of work tests deliberate manipulation and its identification, the present study examines citation and exposure distributions that arise naturally under controlled query conditions. Together, the two lines of inquiry delineate “natural selection” and “strategic intervention” as distinct classes of questions in GEO research.

## Citation and Source Attribution in Large Language Models

Retrieval-augmented generation (RAG) is defined as involving “models which combine pre-trained parametric and non-parametric memory for language generation” (Lewis et al., 2020, p. 9459). Its theoretical significance lies in introducing externally verifiable evidence into large-language-model answers. Such evidence is generally recalled from an external corpus through dense retrieval. Karpukhin et al. (2020) showed that “retrieval can be practically implemented using dense representations alone, where embeddings are learned from a small number of questions and passages by a simple dual-encoder framework” (Karpukhin et al., 2020, p. 6769). The distinction between parametric memory and externally retrieved evidence provides a conceptual basis for examining whether the content of an answer can be traced to the citations returned at that time. This study defines an answer entity for which no corresponding item can be found in the contemporaneous citation pool as an “unmatched exposure,” thereby recording the correspondence between entity exposure and the citation text visible at that time. RAG also raises questions of attribution and faithfulness—whether an answer faithfully uses the sources it cites. Research in this area examines agreement between answers and evidence, citation correctness, and traceability. Liu et al. (2023) treat verifiability as a prerequisite for trustworthy generative search: “A prerequisite trait of a trustworthy generative search engine is verifiability” (Liu et al., 2023, p. 7001). This research tradition corresponds to the citation-absorption categories used here: merely listing a source does not establish that its content has been substantively used in the answer, making it necessary to distinguish nominal, general, and deep citations. Recent audits have also identified quality concerns in the cited sources themselves. Auditing four generative search engines, Allaham and Diakopoulos (2026) found that approximately 16% of cited sources were AI-generated, reporting “evidence of AI-generated sources being cited across all four generative search engines.” This finding indicates that citation status and source quality are distinct measurement dimensions and supports the concurrent recording of source domain, site type, and the 5118-Baidu Composite Quality Score. The present study treats degree of citation absorption, entity exposure, and unmatched status as its principal outcome variables and examines their relationships with source attributes, query design, and platform interface.

## Citation Behavior and the Chinese Platform Ecosystem

Existing studies provide observational evidence across platforms and in the Chinese context regarding the composition of sources cited by search and question-answering systems, the extent of evidentiary support, and user adoption. In an audit of four generative search engines, Liu et al. (2023) found that “a mere 51.5% of generated sentences are fully supported by citations and only 74.5% of citations support their associated sentence” (Liu et al., 2023, p. 7001). This finding demonstrates that citation count alone is insufficient evidence that an answer is adequately supported. The present study therefore distinguishes between a citation being listed and the extent to which source content is actually absorbed. In the Chinese context, Shang et al. (2026) examined the effect of generative AI on the use of traditional search engines and found that content accuracy and multimodal interaction were important features increasing users' willingness to adopt these systems, with user trust and content quality serving as key mediating pathways. That study demonstrates the importance of content quality and trust from the perspective of user adoption; the present study examines the relationships among source type, degree of citation absorption, and entity exposure from the system-output perspective. From a critical perspective, Fang and Ye (2026) discuss the dispersion of knowledge positions in generative search engines as “organized disorder” and argue that the probabilization of semantics and fragmentation of relational structures may create risks for knowledge acquisition and governance. This discussion suggests that evaluating generated answers entails not only determining whether an individual source is correct but also examining how source sets are organized, whether different platform interfaces are stable, and whether highly sensitive domains exhibit different source structures. Existing Chinese-language research has rarely compared Doubao, DeepSeek, Tencent Yuanbao, and Qwen simultaneously using unified citation-level fields; it has also seldom quantified article recency, semantic role, brand and contact-information exposure, and App–Web source-set consistency. The present study establishes unified measurement rules for these dimensions, supplementing the empirical evidence on citation behavior and entity exposure across multiple Chinese platforms.

# Chapter 1 Dataset and Unified Construct Framework

To ensure that subsequent analyses are comparable under a unified measurement framework, this chapter describes the data sources and collection design, analytical granularity and units of analysis, and the operational definitions of the principal constructs and variables. All sample sizes, fields, outcome variables, and interpretive boundaries used in subsequent chapters follow the definitions established here.

## 1.1 Data Sources and Experimental Design

The research corpus was drawn from a controlled data-collection experiment involving mainstream Chinese AI question-answering assistants. Focusing on the visibility of brands and content providers in generative engines, the experiment submitted the same set of queries to the Web and mobile App interfaces of four large-language-model platforms—Doubao, DeepSeek, Tencent Yuanbao, and Qwen—yielding eight platform interfaces in total. Each query–platform-interface combination was collected in three fixed replications. The collection captured the body text of each model answer and detailed information for every source it cited, including the title and body text of the cited webpage, publication time, source domain, citation sequence number, and the model's classification of the citation's semantic similarity and role.

**Table 1-1. Mapping of Platform-Interface Codes (`platform_code`)**

| Large-Language-Model Platform | Web Interface Code | App Interface Code | Provider and Ecosystem |
|---|---|---|---|
| Doubao | DB | DOUBA | ByteDance |
| DeepSeek | DP | DPA | DeepSeek |
| Tencent Yuanbao | TXYB | TXYBA | Tencent |
| Qwen | TYQW | TYQWA | Alibaba |

Note: Some upstream tables—specifically, answer-level Table 1—also contain records for platforms such as BDAI, DYAI, KIMI, and WXYY. Because these records have no corresponding details in the citation-level master table, they are excluded from all analyses in this study.

### 1.1.1 Query Design and Query Space

The query set was designed in a controlled manner along query-formulation dimensions to compare differences in citation and exposure associated with different query intents within the same experiment. The query space was formed by crossing three dimensions:

Query layer (`layer`), comprising six categories: query attribute, extreme and real-world scenarios, industry dimension, prompting style, time sensitivity, and trigger intensity.

Fine-grained topic, comprising 31 `layer–subcat` design combinations, including comparison, reputation, purchase/e-commerce, factual, source-request, and expert-role categories, as well as high, medium, and low levels of time sensitivity and trigger intensity. In principle, each design combination contained approximately 20 queries. Because the labels “high,” “medium,” and “low” are reused across different layers, deduplicating only the `subcat` strings yields 28 textual labels. All references to 31 categories in this study denote the `layer–subcat` design combinations.

Industry dimension, comprising 21 industries, including food and dining, home appliances and digital products, beauty and medical aesthetics, hotels and tourism, healthcare, finance, law, and maternal and childcare services. These industries span both consumer and YMYL domains; YMYL refers to highly sensitive domains—such as healthcare, law, and finance—that directly affect users' vital interests.

The full experiment included 614 unique `prompt_id` values, cross-designed across query layers, 31 `layer–subcat` design combinations, and 21 industry dimensions, covering both consumer and YMYL domains.

### 1.1.2 Primary Key and Data Versioning

The citation-level master table uses `row_uid` as its primary key, constructed as follows: `row_uid = md5(prompt_id | platform_code | quote_index | quote_url | run)[:16]`. This key is unique within the cleaned citation-level master table, ensuring that each cited source can be uniquely located for a given query, platform interface, and collection run; it therefore serves as the anchor for cross-dimensional joins and deduplication. The research design fixed three collection replications, with the `run` field taking values of 1, 2, or 3, and expanded the data by individual cited article. Each chapter selects the subset relevant to its research question. For example, the position analysis includes only platforms that output inline labels, whereas the article-recency analysis includes only records with parsable publication dates. Each table reports its cleaning rules and effective sample size separately.

### 1.1.3 Data Collection and Multimodal Processing Pipeline

Construction of the corpus involved a substantial collection, crawling, and multimodal-normalization pipeline. Its complexity arose primarily from platform heterogeneity, the large number of variables, and rich-media content:

For answer and citation collection, Aidso, an AI-search monitoring tool, was used to monitor queries submitted to the eight platform interfaces and to capture the body text of model answers together with item-level citation details, including title, body text, publication time, source domain, and citation sequence number.

For original-text crawling, the complete body text of every cited webpage was crawled as the source text of the cited article. Interactive and structured components—such as page buttons, pop-up windows, footer contact cards, and JSON-LD—were excluded.

For multimodal conversion of rich media into text, cited and exposed content included both video and image materials. MP4 video was first converted to MP3 audio using `ffmpeg` and then transcribed through automatic speech recognition using Volcengine's One-Sentence Recognition service. For images, such as illustrated cards and screenshots of rankings, a multimodal large language model was prompted to recognize text in the image and extract it. Automatic speech recognition and vision-language-model technologies have matured rapidly in recent years (Radford et al., 2022; Bai et al., 2023), providing a methodological basis for mapping rich media into a unified textual representation suitable for analysis.

The raw data comprised 214,119 records. After field linkage, deduplication, and removal of invalid records, a citation-level master table of 160,860 records and 28 fields was constructed for analysis. Each query was collected in three replications for every platform-interface combination. Each topic-specific analysis then reports the cleaned effective sample size according to variable availability and applicability conditions.

## 1.2 Variables and Field Dictionary

The fields are organized into four groups: identifiers and query design; cited sources and text; citation classification; and contact information and brands. The field dictionary describes the scope and statistical definitions of fields in the internal analytical tables.

**Table 1-2. Field Dictionary I: Identifiers, Query Design, and Cited-Source Fields**

| Field | Type | Values / Range | Meaning and Operational Definition |
|---|---|---|---|
| row_uid | Text | Unique across the full table | Citation-level primary key generated from the query, platform interface, citation sequence number, URL, `run`, and related information; used for joins and deduplication. |
| pk | Text | Unique across the full table | Upstream record primary key used to trace record provenance before and after cleaning. |
| prompt_id | Text / Integer | 614 values | Unique query identifier and primary key for query-level grouping and cross-table joins. |
| platform_code | Categorical | 8 interfaces | Codes for the Web and App interfaces of four platforms. |
| quote_index | Integer | ≥1 | Sequence number of the cited source within the corresponding answer. |
| quote_url | Text | — | URL of the cited webpage. |
| prompt | Text | — | Actual query text submitted to the model. |
| layer | Categorical | 6 categories | Top-level category in the query design. |
| subcat | Categorical | 31 design combinations | Thirty-one `layer–subcat` combinations; deduplicating the text labels alone yields 28 labels. |
| site_name | Text | — | Name of the cited source site. |
| domain | Text | — | Domain of the cited source. |
| published_at | Date / Text | Approximately 67% non-missing | Raw field for the webpage publication time. |
| published_date | Date | — | Normalized date field derived from `published_at`. |
| quote_title | Text | — | Title of the cited webpage. |

**Table 1-3. Field Dictionary II: Citation Classification, Contact Information, Brand, and Text Fields**

| Field | Type | Values / Range | Meaning and Operational Definition |
|---|---|---|---|
| quote_word_count | Integer | ≥0 | Character count of the cited content. |
| Quote_qwx_score | Continuous | 0–1 | Composite SEO indicator constructed from Baidu indexed-page count, 5118 Weight, website-filing age, and filing type. |
| quote_context | Long text | Internal field | Body text of the cited webpage. |
| context | Long text | — | Body text of the model answer. |
| Context_lxfs | Text | — | Contact information identified in the body of the model answer. |
| Quote_lxfs | Text | — | Contact information identified in the body text of the cited source. |
| quote_role | Multilabel | Roles and role combinations | Semantic role performed by the cited content. |
| Context_quote_type | Text | Nominal, general, deep | Raw label for citation-absorption type. |
| matched_brands | Text | — | Set of brands originally matched in the cited text. |

| Field | Type | Values / Range | Meaning and Operational Definition |
|---|---|---|---|
| Context_quote_score_mean | Continuous | 0–1 | Mean semantic-similarity or absorption score for the relationship between the answer and the citation. |
| run | Integer | 1、2、3 | Sequence number for the three repeated collections of the same query–platform-interface combination. |
| quote_role_n | Integer | ≥0 | Number of semantic roles contained in the citation. |
| matched_brands_cleaned | Text | — | Cleaned version of the brand-matching results for the cited text. |
| quote_context_count_zi | Integer | ≥0 | Number of non-whitespace characters in `quote_context`. |

The four components of the 5118-Baidu Composite Quality Score are combined using a fixed formula. Three components—the 5118 Weight Score, website-filing age score, and filing-type score—were obtained through the 5118 Marketing Big Data Platform interface. The fourth, Baidu Webmaster indexed-page count, was obtained through the Baidu Webmaster Platform interface. Both sources are based on actual marketing and search data. The 5118 Marketing Big Data Platform, Baidu Webmaster, and Baidu Index have been used in prior studies as data sources for keyword collection and search-behavior analysis (Yuan, 2022; Han, 2024). The composite score is therefore an operational proxy for a website's traditional SEO performance and is conceptually distinct from institutional credibility.

## 1.3 Unified Construct Framework

Although the analyses appear to address different questions—profiles of cited sources, citation depth, semantic role, citation position, timeliness, brand exposure, contact-information exposure, and cross-interface consistency—they are grounded in the same set of underlying constructs. This section defines these constructs in one place for direct reference in subsequent chapters. Table 1-4 presents each construct, its corresponding fields, and its operational definition.

**Table 1-4. Core Constructs Used throughout the Study and Their Operational Definitions**

| Construct | Field | Operational Definition |
|---|---|---|
| Semantic Similarity Score and Semantic Fit | Context_quote_score | LLM assessment of semantic similarity or citation strength between an answer and a citation (0–1), divided into three levels using thresholds of 0.30 and 0.65. |
| Degree of Citation Absorption | Context_quote_type | Nominal, general, and deep, represented by ordinal labels 1, 2, and 3, respectively. The labels were generated by a large language model under a unified scoring rubric and constitute an operational measure rather than a human-annotated gold standard. |
| Semantic Role | quote_role | Seven multilabel categories—nominal, background, data, definition, procedure, comparison, and other—expanded into long format using the vertical bar as the delimiter. |

| Construct | Field | Operational Definition |
|---|---|---|
| 5118-Baidu Composite Quality Score | Quote_qwx_score | A composite domain-quality score (0–1) produced by the third-party commercial platform 5118, reflecting traffic and SEO influence rather than institutional credibility. The formula is 0.4 × Baidu Webmaster indexed-page count + 0.3 × 5118 Weight Score + 0.2 × website-filing age score + 0.1 × filing-type score; see the provider's technical documentation for details. |
| Institutional Credibility | Derived from website type | Shares of three source categories—government institutions, news media, and encyclopedic knowledge—treated separately from the 5118-Baidu Composite Quality Score. |
| Citation Position | `quote_index` and `[citation:N]` | Citation sequence number within the answer; for platforms with inline labels, the character position of the first occurrence in the body text and its relative position are used. |
| Article Recency | `published_at` → `freshness_days` (derived) | Number of days between the webpage publication date and the collection reference date. A smaller value indicates a more recent article; a larger value indicates lower recency. |
| Cross-Source Redundancy | Derived | Number of distinct citations within the same answer that match a given brand or entity. |
| Exposure | `Context_lxfs`, answer brands, `rich_media` | Whether contact information or a brand is written in the answer body, or whether a shopping card is triggered; an answer-level outcome. |
| Exposure Unmatched to Contemporaneous Citations | Set operation A − P | An entity exposed in an answer but not matched to the contemporaneous citation pool, denoted by A − P. |

### 1.3.1 Two Association Signals: Content Fit and Article Recency

Across multiple analyses, the 5118-Baidu Composite Quality Score was not the leading predictor. Its relative importance in the citation-depth model was approximately 4.5%; its Spearman correlation with publication-age distance was 0.069; and its correlation in the semantic-role analysis was 0.077. Content fit, cross-source occurrence, and citation depth had greater predictive importance. Article recency was only weakly correlated with the 5118-Baidu Composite Quality Score, providing a complementary association signal distinct from that composite measure.

### 1.3.2 Constructive Dependence and the Principle of Feature Exclusion

Several constructs exhibit constructive dependence. Citation depth is mechanically divided from the semantic similarity score using fixed thresholds, making the two nearly synonymous, while the nominal semantic role is nearly identical to a nominal citation at citation-depth level 1. Including such outcome-derived or definition-derived variables in a model alongside the dependent variable produces artificially inflated explanatory power: in this case, $R^2$ can reach 0.97, with 95.8% attributable to these two columns alone. The following uniform principle is therefore applied throughout the study:

For every attribution analysis in which citation or exposure is the dependent variable, variables constructed from the same source as the dependent variable are excluded. Only external, operationally actionable factors—such as query formulation, platform, source, time, structure, and redundancy—are included. Discriminative performance (AUC) and feature-importance ranking, rather than a high $R^2$, are used as the evaluation criteria.

### 1.3.3 LLM-Assisted Annotation and Consistency Control

The semantic similarity score, degree of citation absorption, and semantic role in Table 1-4 were generated item by item by Doubao-Seed-2.1-Pro (high) under a fixed scoring rubric. They therefore constitute model-assisted operational measurements rather than human-annotated ground truth. Prior research suggests that large language models can serve as auxiliary annotators for some text-classification tasks, although their performance varies by task, model, and prompting method (Ziems et al., 2024). This study used a fixed five-dimensional scoring rubric and two prespecified thresholds, 0.30 and 0.65, to improve consistency and auditability of the annotations.

### 1.3.4 Literature Basis for Variable Selection

The core variables in the experimental design were selected on the basis of prior research and supplemented with other available variables. The semantic similarity score captures textual correspondence among the query, citation, and answer (Liu et al., 2023; Zhang et al., 2026). Degree of citation absorption distinguishes the extent to which source content is used in an answer and aligns with the concern for verifiability in research on citation-enabled generation (Gao et al., 2023). Semantic roles cover extractable evidence types such as definitions, data, comparisons, and procedures (Zhang et al., 2026). Article recency corresponds to the updating of time-sensitive knowledge (Vu et al., 2024). Citation sequence number represents positional information in long contexts (Liu et al., 2024). Source attributes and brand frequency support comparisons of sample distributions across sources and entities (Aggarwal et al., 2024; Alipour et al., 2026). Cross-source occurrence count is an additional measurement dimension introduced into this variable system to test the relationship between repeated mentions across sources and citation or exposure outcomes. These studies provide the basis for the operationalization and statistical comparison of the variables.

## 1.4 Unified Analytical Methods

All analyses share a common statistical toolkit. The methods are listed here so that subsequent chapters need report only parameter differences rather than repeat the methodological details.

Feature-importance ranking uses LightGBM gradient-boosted trees (Ke et al., 2017), which natively support categorical variables and nonlinear interactions. Feature importance is cross-validated using both gain and the mean absolute SHAP value (Lundberg & Lee, 2017). In the tree-based predictive models, high-cardinality categories, including domains and site names, are processed using five-fold out-of-fold target encoding (Micci-Barreca, 2001) to prevent data leakage. Correlational analyses use Spearman rank correlation because most dependent variables are zero-inflated, nonnormal, and ordinal; $\rho$ and p values are reported. Associations between categorical variables are evaluated using the chi-square test of independence and Cramér's V, with magnitudes classified as weak (0.1–0.3), moderate (0.3–0.5), or strong (>0.5), supplemented by adjusted standardized residuals and lift. For multivariable attribution, logistic-regression odds ratios indicate direction and magnitude, while OLS and ordinal regression decompose net contributions.

Article recency is represented by the number of days, d, between the webpage publication date and the collection reference date; a smaller d denotes a more recent article. The decay model fits the relative frequency of cited pages across publication-age distances using the log-linear model $N(d) \propto e^{(-\lambda \cdot d)}$ and summarizes the recency-decay scale using the half-life $\ln 2/\lambda$. The half-life concept is adapted from research on literature aging (Burton & Kebler, 1960).

This study classifies degree of citation absorption into nominal, general, and deep categories. A nominal citation indicates that a source is listed but that its content is used to only a limited identifiable extent in the answer. Nominal citations account for more than half of the full dataset, demonstrating that "listing a source" and "substantively using source content" are not equivalent indicators. The label therefore operationalizes comparisons of the extent of content use across citations.

Robustness procedures include random-seed stability tests, for which the feature-importance rankings achieved a cross-seed Spearman consistency of 0.991, together with triangulation across methods, univariate dose–response analysis, transparent reporting of cross-platform consistency, and a dedicated test for crawl truncation.

### 1.4.1 Model Specifications and Statistical Conventions

To make the analytical process auditable, all predictive models were trained with random seed 42. The principal LightGBM hyperparameters were `learning_rate=0.05`, `num_leaves=63` (48 for the brand-selection model and 40 for the unmatched-exposure classification model), `n_estimators=400–700`, `subsample=0.8`, and `colsample_bytree=0.8`. Training and test sets were stratified at either 8:2 or 75:25; high-cardinality categories used five-fold out-of-fold target encoding; and the SHAP sampling seed was 1. Logistic regression and OLS used maximum-likelihood estimation and least-squares estimation, respectively. Cross-interface consistency and source ecology were analyzed using set-based metrics and descriptive statistics. The random forest for contact-information exposure used `n_estimators=100`, `max_depth=8`, and random seed 42 and was evaluated using five-fold `StratifiedKFold`. The 95% confidence interval for AUC was estimated from 100 bootstrap samples, and permutation importance was repeated 20 times. The multivariable logistic regressions estimating exposure odds ratios controlled for query topic, query category, semantic role, source domain, platform, citation character count, and publication-age distance. Because positive cases were sparse in some models, both point estimates and confidence intervals are reported to convey estimation precision. The shopping-card model describes the distribution of platform-native commercial components and differences across platforms.

For statistical inference, proportion-based indicators report binomial confidence intervals, including overall and platform-specific contact-information exposure rates. Platform-level mean Jaccard coefficients for cross-interface consistency report 95% confidence intervals. Between-group differences are measured using the Mann–Whitney U test, the Kruskal–Wallis test, and Cliff's δ effect size; categorical associations are assessed using chi-square p values and Cramér's V. Some odds ratios and half-lives are point estimates without intervals; conclusions involving these values are therefore expressed only in terms of direction and order of magnitude to avoid overinterpreting a single point estimate.

## 1.5 Analytical Design and Statistical Objectives

The analytical design is organized according to how the variables were generated. Query formulations were constructed from prespecified `layer–subcat` combinations and are used for controlled between-group comparisons. Source attributes, structural features, and platform characteristics were observed from platform output and are used to estimate conditional associations and predictive contributions. Table 1-5 summarizes the variable source, comparative design, and statistical objective for each class of factors.

Table 1-5. Variable Sources, Comparative Designs, and Statistical Objectives by Factor Class

| Factor Class | Variable Source | Comparative Design | Statistical Objective |
|---|---|---|---|
| Query Design (`layer` and `subcat`) | Prespecified by the researchers; in principle, approximately 20 queries per design combination. | Controlled between-group comparison by design combination. | Estimate differences among query-formulation categories within the current query set. |

| Factor Class | Variable Source | Comparative Design | Statistical Objective |
|---|---|---|---|
| Source Attributes (5118-Baidu Composite Quality Score, Article Recency, and Domain) | Observed attributes of sources returned by the platforms. | Conditional comparisons in multivariable models. | Estimate conditional associations between source attributes and outcome variables. |
| Structural Features (Cross-Source Occurrence Count, Within-Answer Position, and Semantic Role) | Observed citation structure or model-assisted annotation results. | Group comparisons and predictive models. | Describe structural differences and compare predictive contributions. |
| Model-Interpretation Metrics (Gain and SHAP) | Calculated from the fitted predictive models. | Compare each feature's contribution to the predicted outcome. | Report predictive importance within the model. |

### 1.5.1 Controlled Between-Group Comparisons by Query Formulation

The research queries were constructed from prespecified combinations of `layer` and `subcat`, with approximately 20 queries in each combination in principle. This study compares outcome differences among query-formulation categories within the current query set. In the multivariable model, brand-related queries had an odds ratio of approximately 89 for contact-information exposure relative to the reference group. After control for other variables, this result indicates that query type is the primary stratification factor for contact-information exposure.

### 1.5.2 Conditional Associations of Source and Structural Features

Models of source attributes and structural features compare the conditional associations and predictive contributions of each variable with citation and exposure outcomes. Conditional associations and predictive contributions are reported separately to reflect the distinct statistical objectives of the models. LightGBM gain and SHAP measure each feature's contribution to model prediction, and their rankings are used to compare predictive contributions across variables.

### 1.5.3 Treatment of Definitional Overlap and Variable Confounding

To avoid target leakage, each predictive model excludes features used directly to construct its dependent variable. Citation-absorption depth (`Context_quote_type`) and semantic role (`quote_role`) are maintained as separate analytical fields: the former measures the extent of content use, whereas the latter records the function performed by cited content. Accordingly, semantic similarity is excluded from models whose dependent variable is citation-absorption depth, while `quote_role` is retained for the separate role-by-depth analysis. This separation prevents the two analytical fields from being treated as interchangeable.

### 1.5.4 Observational Levels and Objects of Comparison

This study primarily observes outcomes after a source has entered the citation list. During answer generation, entities that enter an answer can be compared with those that do not within the same answer, and sources cited in the body can be compared with listed sources that do not enter the body, making it possible to identify secondary selection during answer generation. The source-selection analysis takes webpages that entered the citation list as its observational population and reports their internal composition and conditional distributions.

## 1.6 Data Quality and Authenticity Checks

Corpus quality control covers three stages: source authenticity, date parsing, and missing-data treatment.

To verify the authenticity of cited sources, DNS resolution was checked for 422 principal cited domains, of which 421 could be resolved. After invalid placeholders (`expired_url`), missing values, the unresolvable domain `zjurl.cn`, IP fragments, and other false or invalid information were removed, all domains included in the analysis passed validity screening.

The raw non-missing rate for the `published_at` field was approximately 67%, but date formats differed across platforms. After unified parsing, article recency could be calculated for approximately 63% of citation records. The article-recency analysis uses records with parsable publication dates and reports parsing coverage and effective sample size by platform.

For missing-data and replication handling, rows missing the target variable were excluded according to the rules of each topic-specific analysis. Missing `published_at` values were retained as a separate unknown category, preserving the sample while allowing the association of missingness itself to be tested. The research design used three fixed collection replications, with the `run` field taking values of 1, 2, or 3. Each topic-specific analysis separately reports the effective sample size after missingness and applicability conditions are applied.

This chapter standardized the data definitions, construct definitions, and analytical methods. Chapters 2–4 examine cited sources and degree of citation absorption, entity exposure in answers, and App–Web source-set consistency, respectively, reporting results at the same data levels and under a common terminology system.

# Chapter 2 Analysis of Sources Cited by Generative Search Engines

This chapter analyzes the composition of cited sources and the extent to which their content is used in answers. Section 2.1 describes the source ecosystem; Section 2.2 analyzes degree of citation absorption; Section 2.3 compares semantic roles; Section 2.4 examines within-answer position and silent citations; Section 2.5 analyzes the article-recency distribution of cited pages; and Section 2.6 summarizes these results. The chapter thus develops a progressive analytical structure from source ecology to temporal distribution.

Citation and deep citation were strongly associated with content fit, cross-source occurrence, and semantic role, whereas the predictive importance of the 5118-Baidu Composite Quality Score ranked relatively low. Article recency was only weakly correlated with the composite score and can therefore be treated as a complementary association signal.

## 2.1 Profile of the Source Ecosystem: Third-Party Content Dominates and Direct Citation of Official Sites Is Rare

Cited domains were counted at the citation level. Of all 160,860 citations, 159,423 (99.1%) could be uniquely classified into the seven website types below and involved 9,384 unique domains. The remaining 1,437 citations (0.9%) were excluded from the type-composition statistics because their domains were missing or their types could not be determined uniquely. Aggregation by website type reveals a clearly media- and platform-oriented structure in the AI citation ecosystem.

**Table 2-1. Website-Type Composition and Mean Semantic Similarity Scores of Cited Sources**

| Website Type | Number of Citations | Share (%) | Unique Domains | Mean 5118-Baidu Composite Quality Score | Mean Semantic Similarity Score |
|---|---|---|---|---|---|
| News Media | 45,877 | 28.8 | 676 | 0.488 | 0.247 |
| Vertical Industry Portals | 38,497 | 24.1 | 810 | 0.601 | 0.295 |
| Social Media and Independent Content Creators | 26,089 | 16.4 | 68 | 0.525 | 0.290 |
| Brand / Corporate Official Sites | 20,510 | 12.9 | 6,589 | 0.262 | 0.196 |
| E-Commerce and Transaction Sites | 12,112 | 7.6 | 285 | 0.473 | 0.191 |
| Encyclopedic Knowledge | 9,147 | 5.7 | 54 | 0.788 | 0.234 |
| Government Institutions | 7,191 | 4.5 | 1,225 | 0.366 | 0.200 |

News media, vertical industry portals, and social media and independent content creators together accounted for approximately 69% of all citations. Brand and corporate official sites accounted for 12.9% and were distributed across 6,589 unique domains, with an HHI of approximately 0.0008. Here, the HHI is calculated from each domain's share of citations (Kvalseth, 2022) and describes source concentration. These results indicate that cited sources in the sample were dominated by third-party content, while citations to corporate official sites were relatively dispersed. This structure constitutes the principal source ecology of Chinese-language generative engines.

Second, platform citation strategies were highly heterogeneous. The Doubao App interface relied most heavily on leading sites: its top domain accounted for 28.6% of citations and its HHI was 0.094, reflecting strong concentration within the Douyin ecosystem. The DeepSeek Web and App interfaces drew on the broadest range of sources: their top domains accounted for only approximately 5% and their HHIs were approximately 0.01. However, they also had the lowest mean 5118-Baidu Composite Quality Scores, at 0.467 and 0.482, respectively. The Qwen App interface had the highest semantic similarity score, 0.407, and used the fewest domains—975—to support 22,000 citations. Source concentration was therefore highly heterogeneous across platforms.

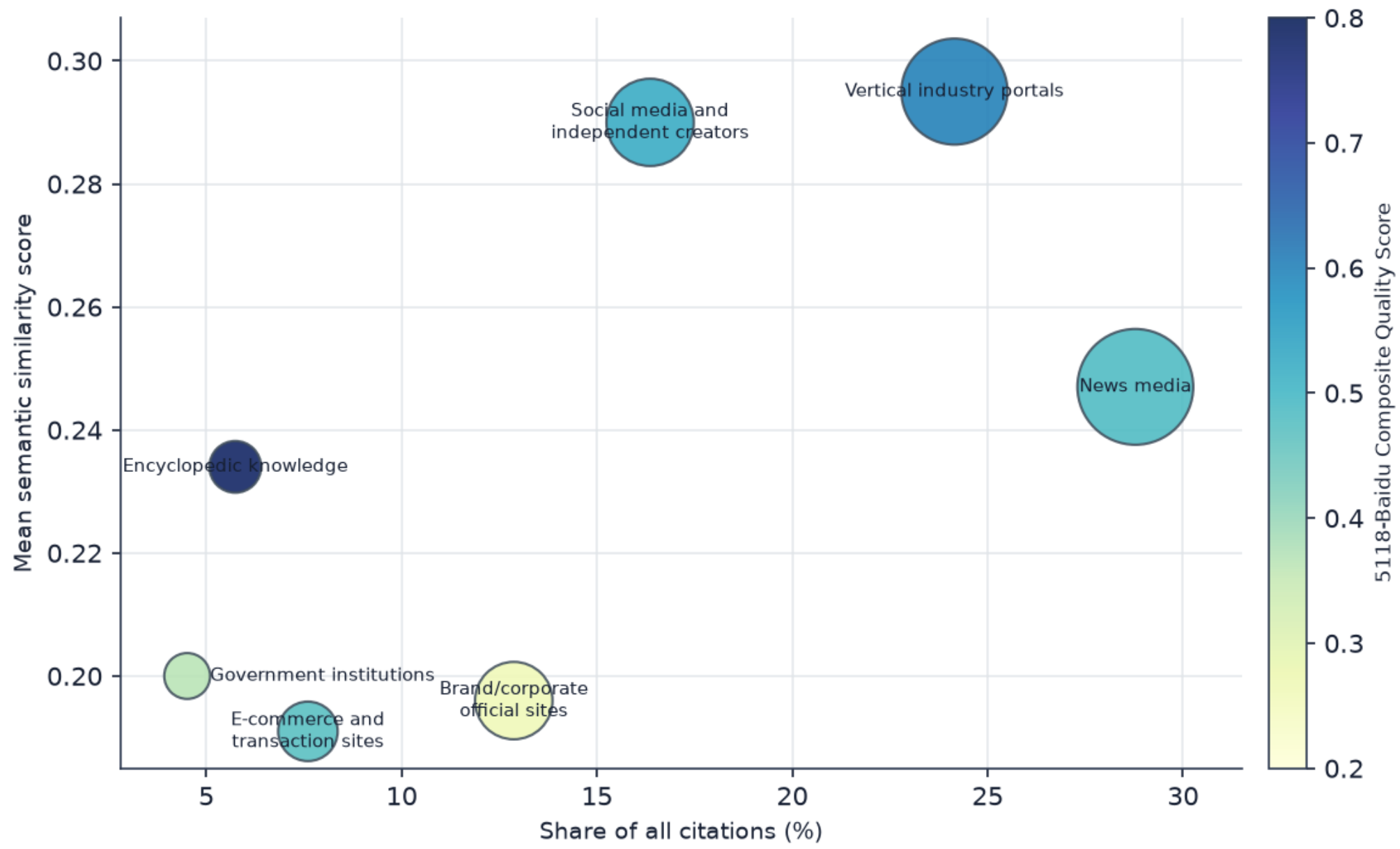

Figure 2-1. Website-Type Composition and Mean Semantic Similarity Scores of Cited Sources

As shown in the figure, the horizontal axis represents each website type's share of all citations, and the vertical axis represents its mean semantic similarity score. Bubble area represents the number of citations, and color represents the 5118-Baidu Composite Quality Score. News media, vertical industry portals, and social media and independent content creators together account for approximately 69% of all citations, while brand and corporate official sites account for 12.9%.

### 2.1.1 Differences in Source Composition between YMYL and Consumer Domains

YMYL domains such as healthcare, law, and finance exhibited source compositions different from those of consumer domains. Government-institution sources accounted for 14.7%, 5.2%, and 8.2% of citations in law, finance, and healthcare, respectively, compared with 1%–4% in most consumer domains. E-commerce and transaction sites accounted for 1.2% and 2.0% in finance and law, respectively, but 21.9% and 10.8% in hotels and leisure. Meanwhile, the mean 5118-Baidu Composite Quality Score was 0.483 for the YMYL group and 0.512 for the consumer group. These differences indicate that institutional source type and the composite SEO indicator measure different attributes and provide complementary measurement dimensions. The domain-level differences further support the discriminant validity between institutional source type and the 5118-Baidu Composite Quality Score.

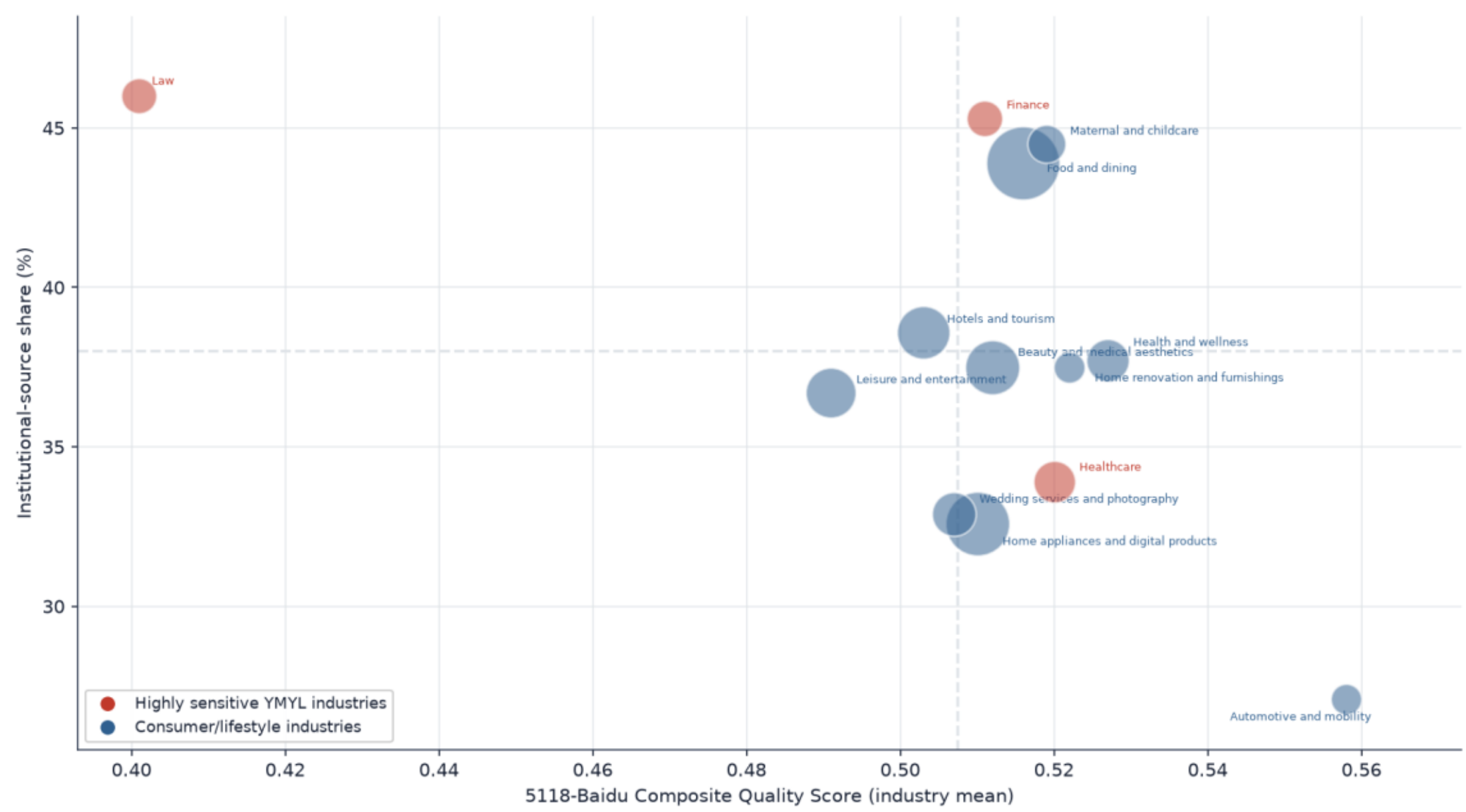

Figure 2-2. Distribution of the 5118-Baidu Composite Quality Score and Institutional-Source Share by Industry

As shown in the figure, the horizontal axis represents the mean 5118-Baidu Composite Quality Score for each industry, and the vertical axis represents the combined share of sources from government institutions, news media, and encyclopedic knowledge sites. Bubble area represents the number of citations; color distinguishes highly sensitive YMYL industries from consumer and lifestyle industries; and the dashed lines indicate the industry means of the two indicators.

### 2.1.2 Query-Level Concentration: Dispersed for Most Queries but Monopolized for a Few

Concentration was measured by the number of citations from the same domain for each query. For approximately 95.3% of queries, the top domain accounted for less than 30% of citations; this share was at least 50% for only 0.8% of queries. Highly concentrated records occurred more often for relatively deterministic queries involving government services, after-sales information, facts, and definitions, such as "How do I apply for public rental housing in Zhengzhou?" and "What is the official after-sales telephone number for XGIMI projectors?" Query-level concentration followed a clear gradient by task type: sources were more concentrated for deterministic queries and more dispersed for open-ended recommendation queries.

## 2.2 Factors Associated with Degree of Citation Absorption: Content Fit, Source Attributes, and Query Characteristics

Traditional search research commonly examines SEO and link-authority signals such as PageRank, inbound links, and domain registration age (Evans, 2007). Accordingly, this section tests associations between source attributes, including the 5118-Baidu Composite Quality Score, and degree of citation absorption. The semantic similarity score, `Context_quote_score`, was divided into three levels using prespecified thresholds: below 0.30 was classified as nominal, 0.30–0.65 as general, and at least 0.65 as deep. The analysis used 160,860 valid citation records. To reduce the risk of circular explanation, the model excluded definition-derived variables constructed from the same source as the dependent variable. Feature importances are shown in Table 2-2.

**Table 2-2. Relative Importance of External Factors Affecting Citation Depth, Measured by LightGBM Gain**

| Factor | Importance (%) | Factor | Importance (%) |
|---|---|---|---|
| Citation Character Count | 23.97 | Number of Brands Matched | 7.69 |
| Site Name of the Cited Source | 15.48 | Total Number of Cited Articles | 5.34 |
| Domain | 8.79 | Citation Sequence Number | 5.22 |
| Query Topic | 8.52 | 5118-Baidu Composite Quality Score | 4.52 |
| Article Recency | 8.17 | Platform | 2.59 |
| Answer-Text Length | 7.78 | — | — |

In the degree-of-citation-absorption model, citation character count, website and domain, and query topic had greater predictive importance than the 5118-Baidu Composite Quality Score, whose relative importance was 4.52%. The deep-citation rate was 10% for citations shorter than 100 characters and 39% for those longer than 1,000 characters. It was 8% when no brands were matched and 44% when more than 10 brands were matched. From the lowest to the highest range of the 5118-Baidu Composite Quality Score, the deep-citation rate changed from approximately 16% to approximately 14%. Taken together, the importance ranking and binned results indicate that this indicator was a secondary predictor in the current model.

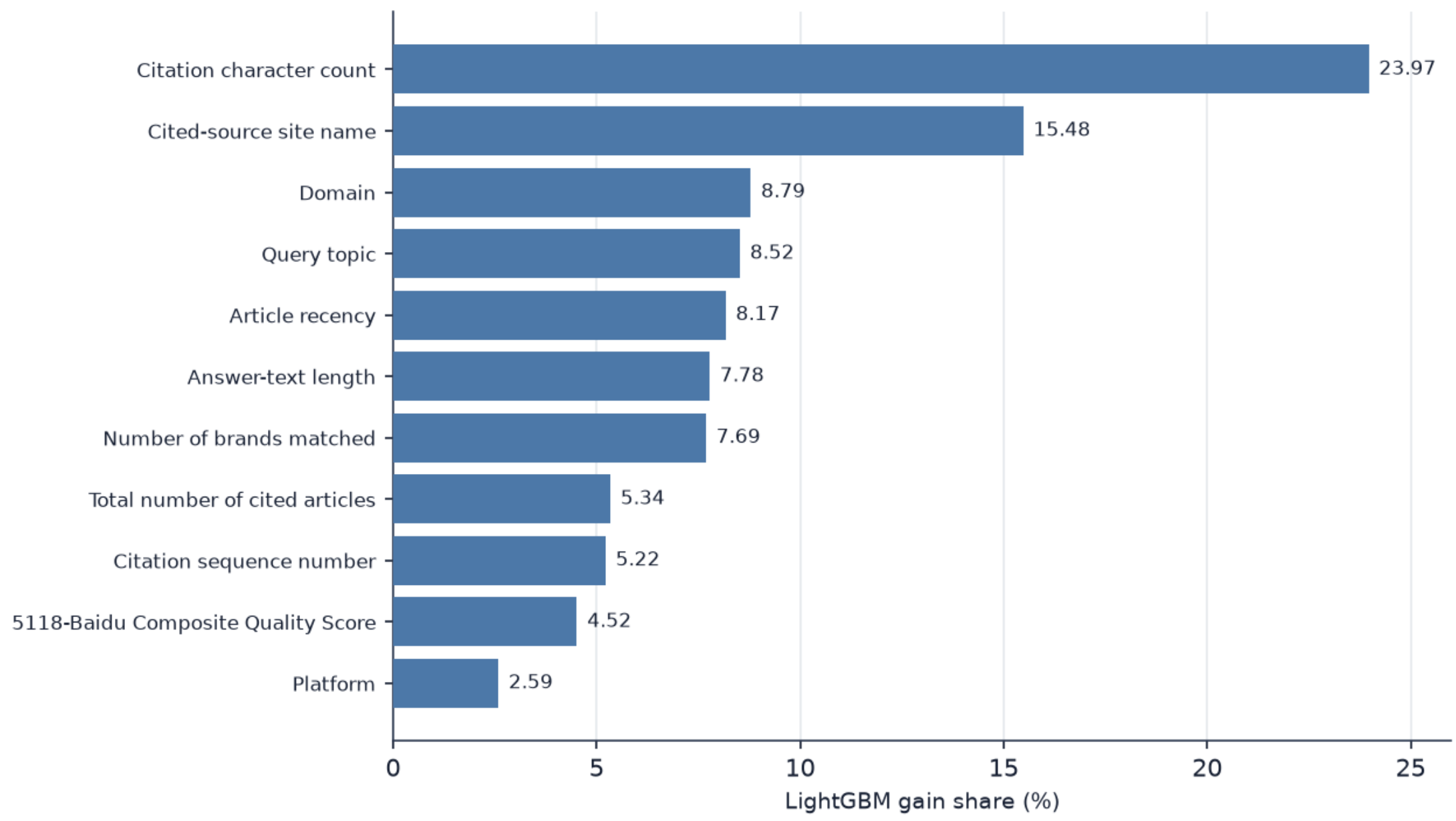

Figure 2-3. Relative Feature Importance in the Citation-Depth Prediction Model (LightGBM Gain)

As shown in the figure, the horizontal axis represents the share of LightGBM gain. The 5118-Baidu Composite Quality Score had a relative importance of 4.52%, ranking 10th among 11 variables.

### 2.2.1 Classification Discrimination and Explanatory Power for Continuous Outcomes

The classification models showed relatively strong discrimination of citation status: the AUC was 0.849 for identifying whether a source was substantively cited and 0.806 for identifying whether it was deeply cited. The regression model using semantic similarity score as the dependent variable had an $R^2$ of 0.36. AUC measures a classification model's ability to distinguish positive from negative cases, whereas $R^2$ measures the proportion of variation in a continuous outcome explained by a regression model; the two metrics correspond to different statistical tasks. Variance decomposition showed that, after controlling for website and query topic, 80.4% of the variance in semantic similarity score remained within groups, indicating that the available structural metadata did not explain all record-level variation.

### 2.2.2 Robustness and Cross-Platform Differences

Three robustness checks support the ranking above. First, retraining with five random seeds yielded a mean Spearman consistency of 0.991 for the importance rankings. Second, descriptive statistics, Spearman correlations, and a tree model with SHAP independently identified the same leading factors. Third, in a dedicated test addressing potentially incomplete citation crawling, removing citation character count reduced discrimination only slightly, from AUC 0.801 to 0.794, leaving the conclusion unchanged. Cross-platform differences were present: the correlation between character count and depth was positive on most platforms, reaching +0.305 for Qwen, but was near zero for Doubao at −0.015. The overall pattern therefore held, although its magnitude varied by platform. The platform-level deep-citation rate was highest for the Qwen App interface at 36.8%, approximately 4.6 times the lowest rate of 8.0% for the Yuanbao App interface.

## 2.3 Semantic Role and Degree of Citation Absorption

The citation semantic-role field, `quote_role`, describes the function performed by a citation in an answer, including definition, data, comparison, and background. This section examines the association between semantic role and degree of citation absorption. `Context_quote_type` is mapped to ordinal scores of 1, 2, and 3, and, following the principal rule established in Chapter 1, the semantic similarity score is excluded because it is constructively related to the dependent variable. The underlying citation table contains 160,860 citation records. Because one citation can perform multiple roles, vertical-bar-delimited roles are expanded to 166,033 role records; the role-level regression uses 165,986 complete cases after records with missing model fields are excluded.

**Table 2-3. Mean Citation Depth and Deep-Citation Share by Semantic Role**

| Semantic Role | Frequency | Share (%) | Mean Citation Depth | Deep Citations (%) | Nominal (%) |
|---|---|---|---|---|---|
| Definition and Explanation | 26,738 | 16.1 | 2.402 | 41.1 | 2.5 |
| Procedural Steps | 6,666 | 4.0 | 2.297 | 36.0 | 7.4 |
| Comparison | 5,691 | 3.4 | 2.203 | 29.7 | 9.8 |
| Data Support | 28,090 | 16.9 | 2.179 | 30.2 | 13.0 |
| Background Context | 28,988 | 17.5 | 1.698 | 6.3 | 35.4 |
| Other | 15,930 | 9.6 | 1.063 | 0.2 | 92.8 |
| Nominal (No Substantive Use) | 53,930 | 32.5 | 1.002 | 0.0 | 99.5 |

Degree of citation absorption differed across semantic roles. Definition and explanation had a mean depth of 2.40 and a deep-citation rate of 41.1%; procedural steps, comparison, and data support also showed relatively high levels. Background context was lower, while the nominal and “other” categories were close to the nominal baseline by definition. The multivariable regression had an $R^2$ of 0.622, and the semantic-role variables made a relatively large predictive contribution. By contrast, the 5118-Baidu Composite Quality Score, article recency, and cited-text length were only weakly correlated with citation depth, with Spearman correlations including 0.077 and −0.015, respectively. These results indicate that semantic role made a relatively large predictive contribution in the current model.

The distribution of semantic roles differed across platforms. The chi-square statistic for the association between role and platform was 13,385 ($p < 0.001$), with Cramér's V = 0.164 (Cox & Holcomb, 2022). The mean depth and deep-citation rate were 1.87 and 22.3% for Doubao, compared with 1.42 and 7.8% for Tencent Yuanbao. Textual features derived from `jieba` part-of-speech tagging and registration of brand terms as out-of-vocabulary items indicated that definition and procedural citations had higher information density and shorter sentence length. These co-occurring patterns suggest that evidence organization is an important textual feature associated with differences in deep citation.

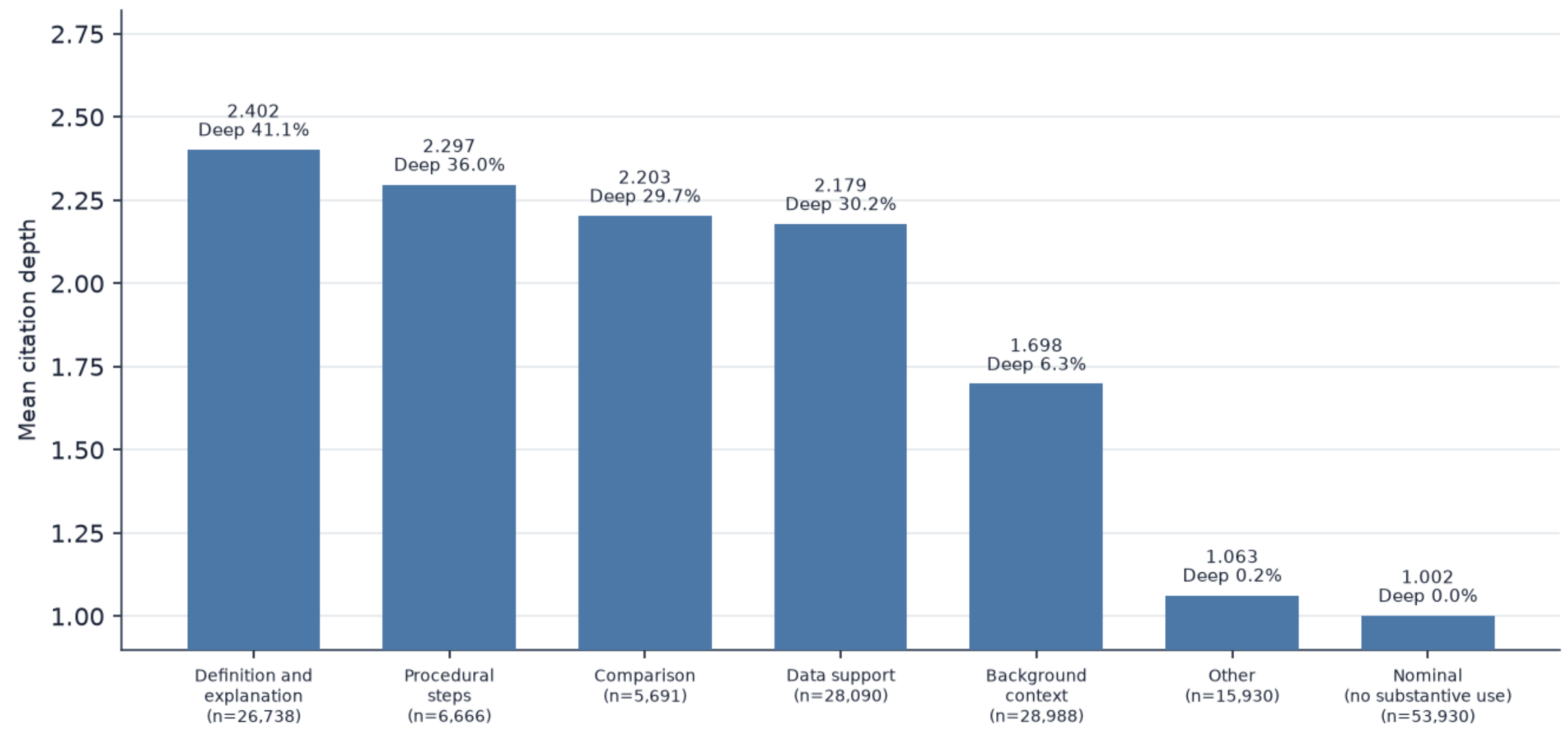

Figure 2-4. Mean Citation Depth and Deep-Citation Share by Semantic Role

As shown in the figure, bar height represents mean citation depth. The first line above each bar reports the mean, and the second reports the deep-citation share; n in parentheses denotes the number of citation records for each semantic role. Nominal citations are assigned a depth of 1, general citations a depth of 2, and deep citations a depth of 3.

## 2.4 Citation Position and Silent Citations: Structural Position Effects and Nearly 40% Never Entering the Answer Body

For the platforms that output inline citation labels in the form `[citation:N]`—the DeepSeek and Tencent Yuanbao interfaces coded DP, DPA, TXYB, and TXYBA—the location of each citation in the answer body can be determined. Doubao and Qwen do not provide such labels. Among 57,558 citations on the labeled platforms, 34,914 had identifiable positions.

**Table 2-4. Antecedents and Presentation Effects of Citation Position (Spearman ρ)**

| Variable Relationship | Spearman ρ | Interpretation of the Correlation |
|---|---|---|
| Order of Appearance and Paragraph Coverage | −0.296 | Weak negative correlation: citations appearing earlier generally cover a larger share of paragraphs. |
| Order of Appearance and Number of In-Text Citations | −0.316 | Weak negative correlation: citations appearing earlier are generally cited more often. |

| Variable Relationship | Spearman ρ | Interpretation of the Correlation |
|---|---|---|
| 5118-Baidu Composite Quality Score and Position | −0.010 (not significant) | No significant correlation was found between the two variables. |
| Semantic Similarity Score and Position | −0.049 | Extremely weak negative correlation: content with higher semantic similarity tends to appear slightly earlier, but the association is limited. |
| Article Recency and Position | −0.007 (not significant) | No significant correlation was found between the two variables. |
| Position and Number of Brands | −0.074 | Extremely weak negative correlation: citations appearing earlier involve slightly more brands. |
| Position and Contact Information | −0.004 (not significant) | No significant correlation was found between position and the presentation of contact information. |

A smaller value for position or order of appearance indicates that content appears earlier or closer to the beginning of the answer. “Not significant” indicates that the correlation did not reach statistical significance ($p \geqslant 0.05$). Citations appearing earlier in an answer generally covered more paragraphs and were cited repeatedly more often. The Spearman correlation between citation position and coverage was approximately −0.30. The first citation accounted, on average, for 25.5% of all inline citations, with an HHI of 0.242. By comparison, the 5118-Baidu Composite Quality Score and article recency were only weakly associated with position. This stable positional gradient makes within-answer order an important structural feature of citation absorption.

Each answer listed an average of 10.5 sources, of which approximately 6.9 entered the answer body as inline citations and 3.5 did not appear in the body. Silent citations accounted for 39.3% of the labeled-platform citation table. The mean semantic similarity scores for silent and inline citations were 0.168 and 0.259, respectively, while their mean 5118-Baidu Composite Quality Scores were 0.486 and 0.501. This comparison indicates that in-answer citation status was more strongly associated with semantic similarity score than with the 5118-Baidu Composite Quality Score. Nominal citations accounted for 55.1% of the full table.

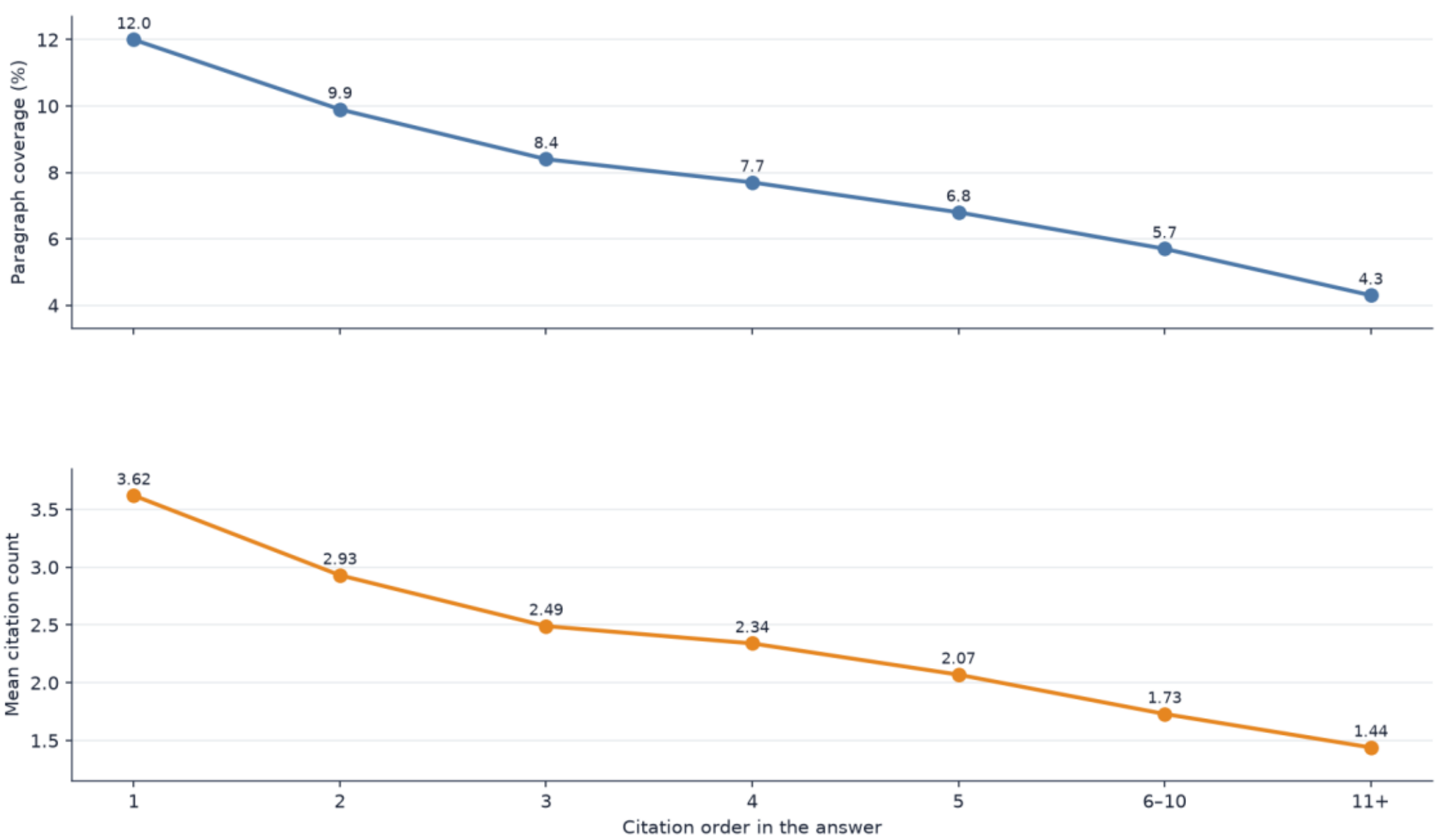

Figure 2-5. Relationships of Citation Order with Paragraph Coverage and Mean Number of In-Text Citations

As shown in the figure, the upper panel presents paragraph coverage and the lower panel presents the mean number of in-text citations. In both panels, the horizontal axis represents the order in which a citation appears in the answer, with 1 indicating the earliest citation. Both indicators decline as citation order moves later, with Spearman correlations of approximately −0.30 and −0.32, respectively.

## 2.5 Article Recency of Cited Pages and Differences by Time Sensitivity

The final citation signal is article recency. This study measures article recency as the number of days between the webpage publication date and the collection reference date, with a smaller number indicating a more recent article. Query time sensitivity is used as the grouping variable. Results are cross-validated under two definitions—a controlled design-axis definition and a regex-based full-dataset definition. The cleaned primary-analysis sample contains 97,968 records, and four subquestions are tested. Between-group differences are measured using the robust ordinal effect size Cliff's δ (Cliff, 1993; effect-size thresholds follow Macbeth et al., 2011), with 95% confidence intervals estimated from 1,000 bootstrap samples. The direction of between-group differences was consistent under the two definitions, with Cliff's δ values of −0.521 and −0.191, supporting the robustness of the article-recency results to the grouping definition.

**Table 2-5. Timeliness Differences, Stage of Effect, and Half-Life**

| Subquestion | Key Result | Effect / Statistic |
|---|---|---|
| H1 Between-Group Difference | The median publication-age distances of pages cited for high-, medium-, and low-timeliness queries were 18, 52, and 101 days, respectively; a smaller value indicates a more recent article. | Cliff's δ=−0.521（95% CI [−0.549, −0.491]）； Kruskal-Wallis p<1e-180 |
| H2 Stage of Effect | Article-recency differences occur in retrieval selection and citation, not in within-answer ranking. | Within-answer position × publication-age distance: ρ = 0.041; deeply cited pages: 73 days < general citations: 87 days. |
| H2 Independence | Article recency is independent of the 5118-Baidu Composite Quality Score. | Publication-age distance × 5118-Baidu Composite Quality Score: ρ = 0.069; publication-age distance × citation absorption: −0.009. |

| Subquestion | Key Result | Effect / Statistic |
|---|---|---|
| H3 Platform Heterogeneity | Doubao and DeepSeek show the strongest recency response; Yuanbao cites the oldest pages. | DeepSeek Web: high timeliness, 55 days vs. low timeliness, 181 days. |
| H4 Half-Life | The relative frequency of article citations decays exponentially with publication-age distance. | High timeliness: 39 days; low timeliness: 68 days; full dataset: 60 days. |

Among cited pages with publication dates, pages corresponding to high-timeliness queries were more recent: 61% were published within 30 days, compared with a full-dataset baseline of 32%, and the between-group effect size was moderate to large. Within-answer position was almost unrelated to publication-age distance, whereas publication-age distributions differed across time-sensitivity groups. The correlation between publication-age distance and the 5118-Baidu Composite Quality Score was 0.069. In within-domain comparisons, pages cited for high-timeliness queries were, on average, approximately 45 days more recent than those cited for low-timeliness queries. Article recency was weakly correlated with the 5118-Baidu Composite Quality Score, and high-timeliness queries showed a clear tendency to cite more recent pages.

Within the high-timeliness query sample, the relative frequency of cited pages decayed approximately exponentially with publication-age distance. The fitted half-life was approximately 39 days: for each additional 39 days of publication-age distance, relative frequency fell to half its preceding level. The fitted half-lives for low-timeliness queries and the full dataset were approximately 68 and 60 days, respectively. Categories with frequent updates—including home appliances and digital products, home renovation, automobiles, and maternal and childcare products—showed larger between-group differences in recency.

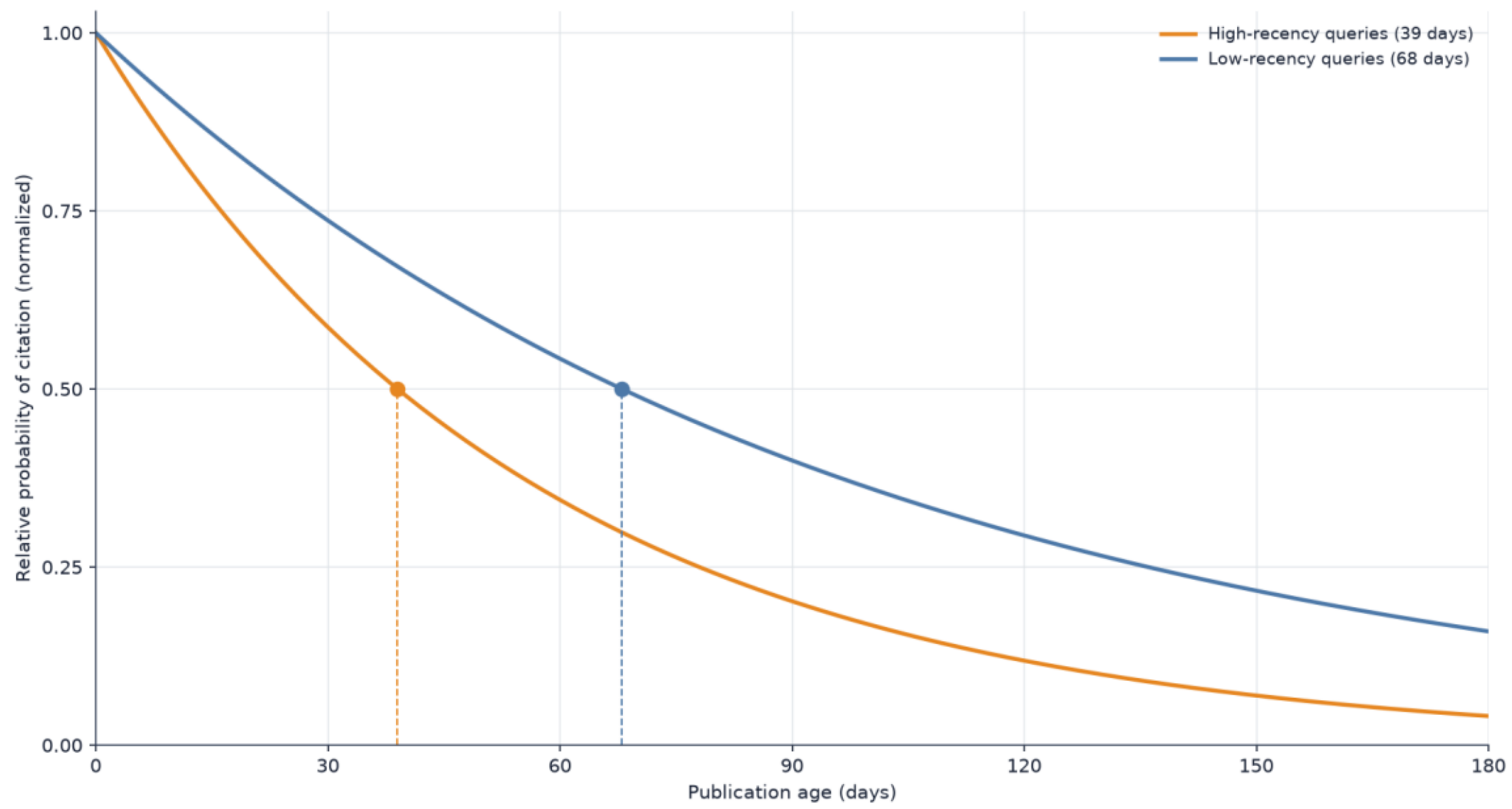

Figure 2-6. Exponential Fits of Relative Frequency over Publication-Age Distance under Different Query Time-Sensitivity Levels

As shown in the figure, the curves are drawn from the exponential model, and the vertical axis represents normalized relative frequency. Dashed lines mark the publication-age distance at which the fitted frequency falls to half its initial level. The fitted scales for high- and low-timeliness queries are approximately 39 and 68 days, respectively. The plotted parameters describe the temporal distribution within the sample of cited pages.

## 2.6 Summary: Principal Associations of Cited Sources and Degree of Citation Absorption

The five analyses in this chapter summarize citation characteristics in the sample along four dimensions: source composition, degree of citation absorption, within-answer position, and article recency.

1. Overall, cited sources were dominated by third-party content, while corporate official-site sources were relatively dispersed. Further differentiation occurred in whether listed sources entered the answer body and the extent to which their content was absorbed into the answer.

2. Across analyses, the predictive importance or correlation of the 5118-Baidu Composite Quality Score was lower than that of the principal variables in the corresponding models: its relative importance was 4.52% in the degree-of-citation-absorption model; its correlation was 0.077 in the semantic-role analysis; its correlation was approximately −0.010 in the within-answer position analysis; and the correlation between publication-age distance and the score was 0.069. Although the four analyses used different dependent variables and models, their conclusions were directionally consistent: the composite score was not the leading predictor in any analysis.

3. Article recency was weakly correlated with the 5118-Baidu Composite Quality Score and showed clear distributional differences across time-sensitivity groups, indicating that the two measures provide different informational dimensions.

4. Source concentration, semantic similarity score, deep-citation rate, and recency response differed across the eight platform interfaces, as shown in Sections 2.1, 2.2, 2.3, and 2.5. Chapter 4 further compares source sets and answer-form differences between the App and Web interfaces of the same platform.

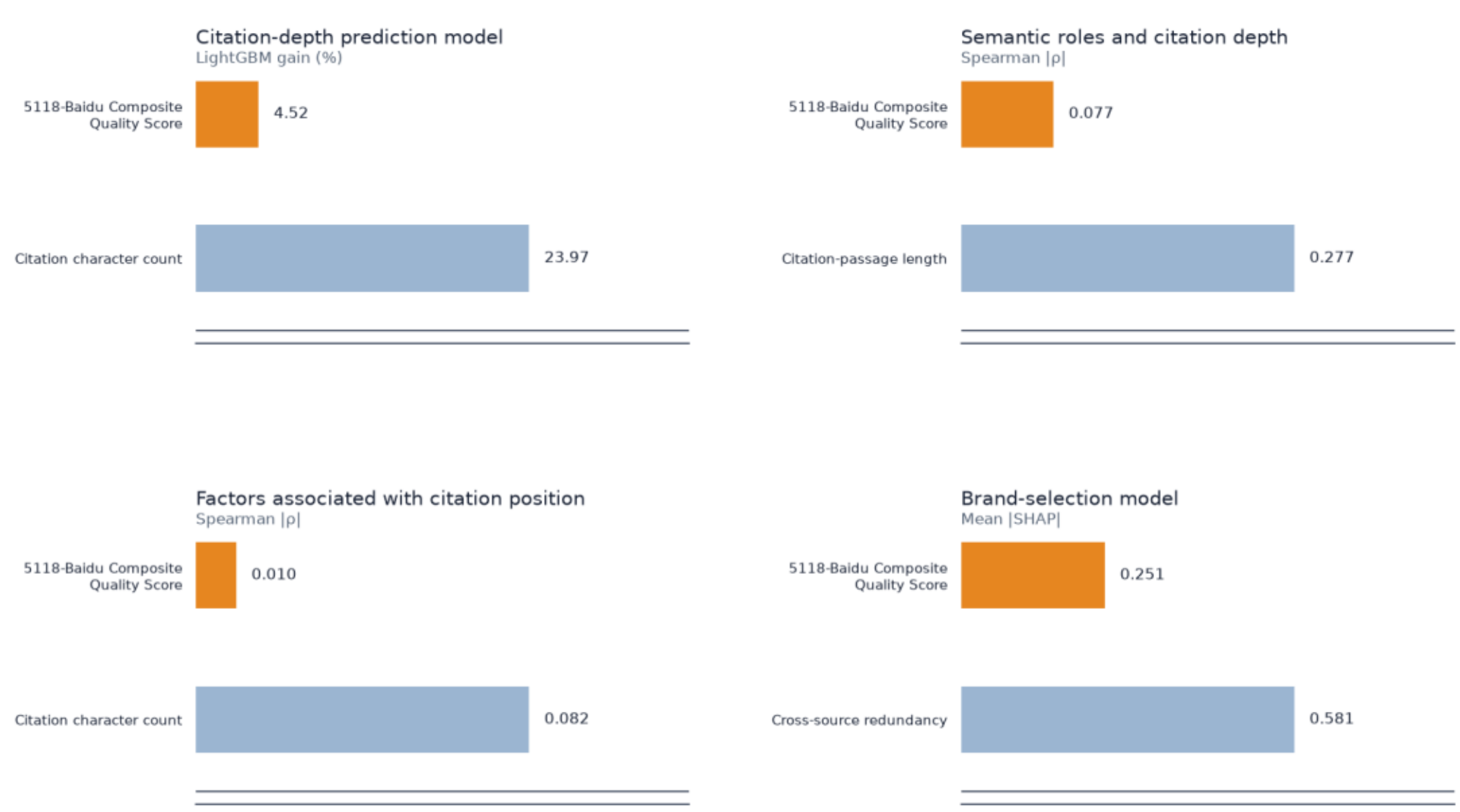

Figure 2-7. Values of the 5118-Baidu Composite Quality Score and Reference Variables across Four Analyses

As shown in the figure, the four panels use LightGBM gain, Spearman |ρ|, and mean |SHAP|, respectively, and retain the original metric scale of each analysis on the horizontal axis. Each panel compares only the 5118-Baidu Composite Quality Score with the leading variable in that analysis.

This chapter described cited sources, degree of citation absorption, within-answer position, and article recency. Chapter 3 further examines whether brands, contact information, and commercial components in cited sources appear in answers and specifies the statistical definition used in each analysis.

# Chapter 3 Analysis of Answer Text from Generative Search Engines

Chapter 2 described cited sources and their degree of citation absorption. This chapter further examines whether entities found in cited sources appear in answers. Three analyses consider the presentation of brands, contact information, and platform-native commercial components. The chapter shows that exposure narrows the set of cited entities further: most cited entities do not enter the answer. Cross-source occurrence, citation depth, and content fit have relatively high predictive importance, whereas the 5118-Baidu Composite Quality Score does not rank first. A further principal finding is that the share of unmatched exposures varies systematically with brand frequency, citation-pool size, and query type.

## 3.1 Two-Layer Brand Sets and Selection Rates: Strong Selectivity after Citation

To compare brands in the citation pool with brands in answers, this study constructed a citation-pool brand set and an answer brand set for each of 4,493 answers—that is, query–platform-interface combinations. Both sets were recalculated from the full text using the same extraction pipeline to reduce discrepancies among abstracts, full text, and dictionary-based definitions. The citation-pool brand set, P, is the union of brands appearing in the body text of all webpages cited by an answer. The exposed-brand set, A, comprises brands appearing in the answer body. This yields three observed states: selection, $P \cap A$, in which a brand appears in the citation pool and is exposed in the answer; elimination, $P - A$, in which a brand appears in the citation pool but is not exposed; and unmatched exposure, $A - P$, in which a brand appears in the answer but cannot be matched to a brand in the contemporaneous citation pool. $A - P$ records the matching status between answer entities and the citation pool available at that time.

**Table 3-1. Three-State Brand Classification and Overall Selection Rate**

| State | Definition | Size / Rate |
|---|---|---|
| Candidate Pool (P) | All answer × in-pool brand candidates | 632,256 candidates |
| Selected (P ∩ A) | Supported by a source and written into the answer | Overall selection rate: 8.3% |
| Eliminated (P − A) | Supported by a source but not exposed | Approximately 91.7% of in-pool brands |
| Unmatched Exposure (A − P) | Exposed in the answer but not matched to the contemporaneous citation pool | Unmatched rate: 13.02% |

Of the brands identified in citation pools, 8.3% also appeared in the answer body, while approximately 91.7% did not enter the answer. These proportions demonstrate that a brand appearing in a cited source is not equivalent to its receiving exposure in the answer and quantify the marked narrowing from citation-pool brands to answer brands.

## 3.2 Factors Associated with Brand Selection: Cross-Source Occurrence Count Has the Highest Predictive Importance

A LightGBM binary classifier was fitted to 632,256 candidates. The positive-class rate was 8.3%, AUC was 0.87, and PR-AUC was 0.51 (Davis & Goadrich, 2006; Saito & Rehmsmeier, 2015). Feature contributions were ranked using mean |SHAP| and their directions were validated using univariate binned selection rates. Results are presented in Table 3-2.

**Table 3-2. Factors Associated with Brand Selection: SHAP Importance and Selection-Rate Gradients**

| Driver | Mean \|SHAP\| | Univariate Selection-Rate Gradient (Low → High) |
|---|---|---|
| Cross-Source Redundancy | 0.581 | Matched in 1 citation: 2.1% → matched in 11+ citations: 59.9% |
| Semantic Similarity Score | 0.364 | Score of 0: 1.1% → score of 0.6–1: 21.1% |
| 5118-Baidu Composite Quality Score | 0.251 | 0–0.3 band: 2.7% → 0.85–1 band: 16.5% |
| Citation Depth | 0.242 | Nominal: 1.5% → general: 7.2% → deep: 20.7% |
| Citation Position | 0.166 | Position 12+: 2.4% → first position: 18.5% |
| Role: Definition | 0.119 | (See Section 3.2.1 for the role dimension.) |

First, cross-source occurrence count had the highest predictive importance in the brand-selection model (SHAP = 0.581). Brand-selection rates were 2.1% when matched in one citation and 59.9% when matched in at least 11 citations, forming a monotonic increase. The 5118-Baidu Composite Quality Score followed a positive gradient but ranked lower, with SHAP = 0.251. Its means in the selected and eliminated groups were 0.66 and 0.53, respectively; this difference was smaller than those for cross-source occurrence count (8.0 vs. 2.3) and semantic similarity score (0.62 vs. 0.30).

The brand-selection rate varied with degree of citation absorption: the selection rates associated with deep and nominal citations were 20.72% and 1.51%, respectively, making the former approximately 13.7 times the latter. After cross-source occurrence count and semantic similarity score were simultaneously included, the odds ratio for degree of citation absorption fell below 1, indicating substantial shared explanatory information among these variables. Citation position was also associated with selection rate: rates were 18.5% for the first position and 2.4% after the 12th position. This positional gradient indicates that brand-selection rate declines as the citation sequence number increases.

Robustness tests showed that the Pearson correlation between cross-source occurrence count and total brand frequency was r = 0.04. After total brand frequency was controlled, the odds ratio changed from 2.71 to 2.87, and selection rates increased monotonically with cross-source occurrence count within low-, medium-, and high-frequency brand groups. The stable gradient between cross-source occurrence count and brand-selection rate therefore remained after total brand frequency was controlled.

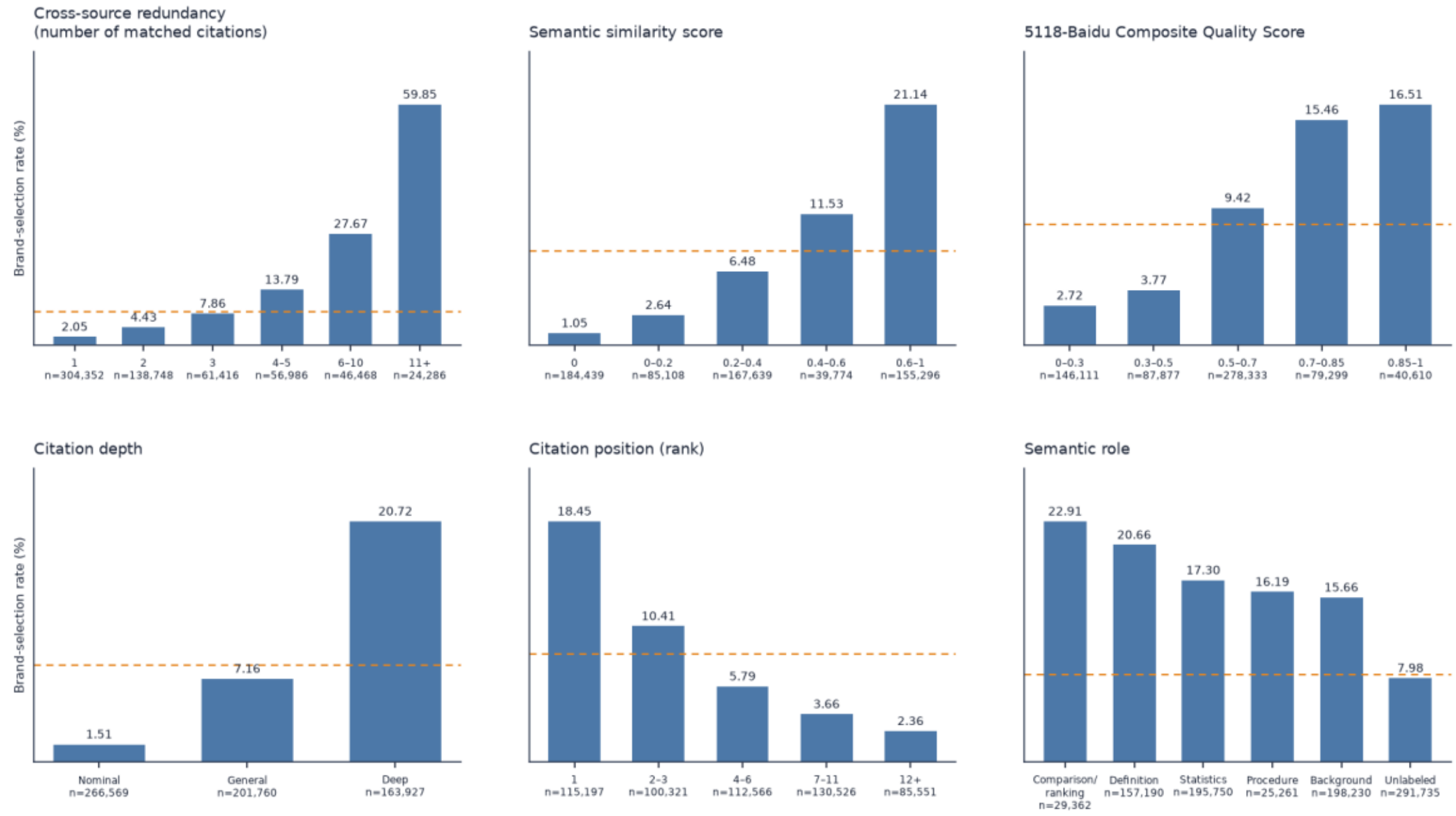

Figure 3-1. Binned Relationships between Brand-Selection Rate and Associated Factors (Dashed Line = Overall Baseline of 8.3%)

As shown in the figure, the vertical axis of each panel represents the brand-selection rate, and the dashed line marks the overall baseline of 8.3%. Selection rates are reported above the bars, and n in parentheses on the horizontal axis denotes the number of analytical instances entering each bin for that analysis.

### 3.2.1 Semantic Roles: Occupying Comparison, Definition, and Data Positions

Marginal selection rates calculated by citation semantic role were 22.9% for comparisons and ranked lists, 20.7% for definitions and concepts, 17.3% for statistics and data, and 16.2% for processes and methods, all above the overall baseline of 8.3%. Records labeled as having no citation had a rate of 7.98%. After the original "other" role was further classified using titles and source domains, comparisons and ranked lists comprised approximately 77,000 instances and had a selection rate of 20.49%. These results indicate that particular semantic roles are associated with higher brand-selection rates. Because secondary classification affects category sizes and proportions, the classification rules and results should be reported together to enable verification.

## 3.3 Brand Exposure Unmatched to Contemporaneous Citations: Approximately One-Tenth of Exposures

After the brand dictionary was reviewed and countries, regions, cities, and generic terms were excluded, the unmatched brand-exposure rate was 13.02%: 7,855 of 60,311 exposure instances could not be matched to the contemporaneous citation pool. Across all exposed-brand instances, 9,240 distinct non-generic brand names were identified.

The classifier for unmatched brand exposure had an AUC of 0.757. Total brand occurrence frequency and the number of brands in the citation pool had relative importances of 34.6% and 25.0%, respectively. Lower brand frequency and a smaller citation pool corresponded to a higher unmatched rate, indicating predictable structural differences in unmatched exposure. By platform, the unmatched rate was 16.9% for DeepSeek Web and 10.1% for Doubao Web. By query formulation, the rates were 27.5% for extremely vague queries, 24.3% for highly time-sensitive queries, and 23.0% for purchase-related local-life queries.

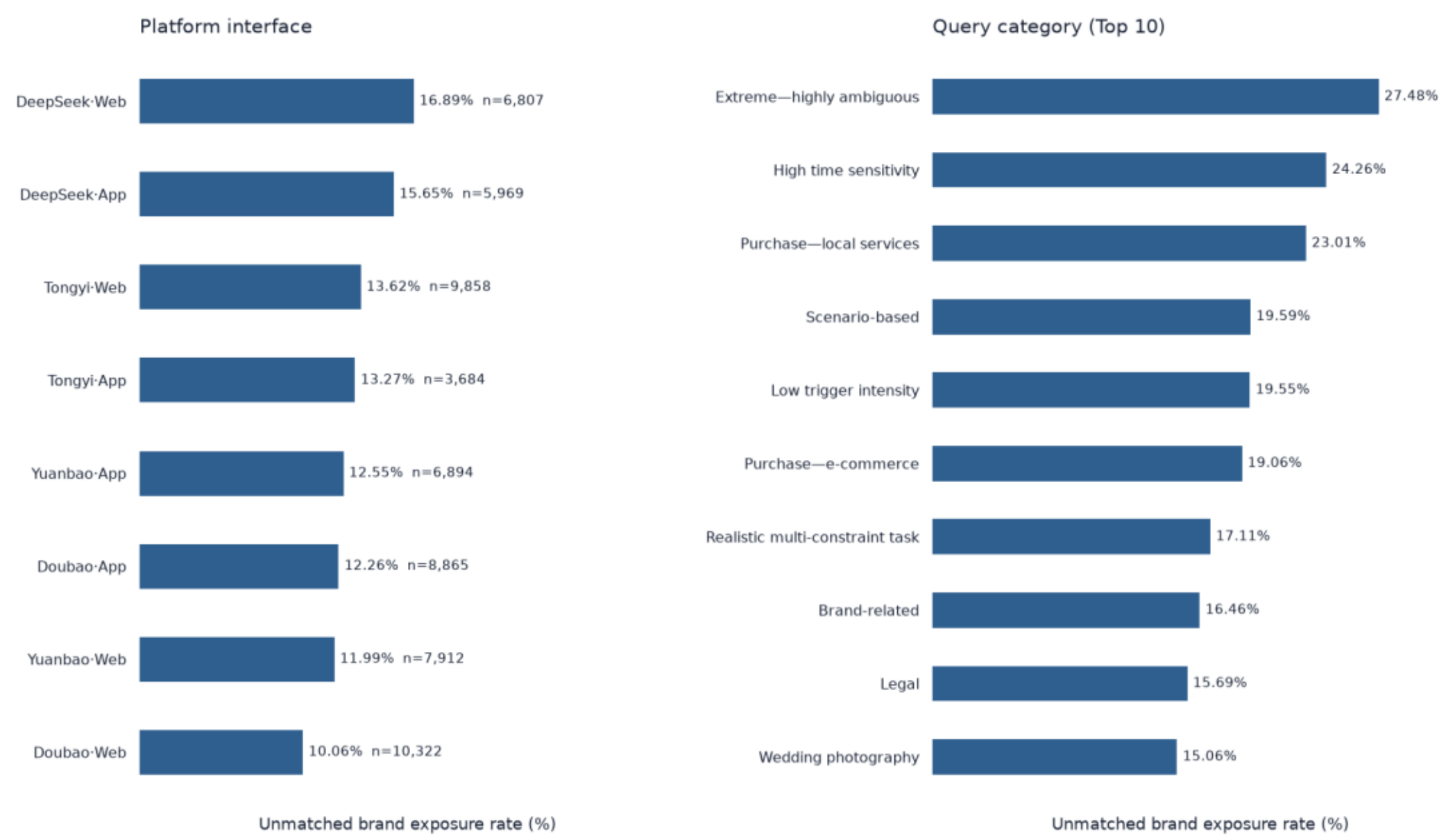

Figure 3-2. Platform and Query-Formulation Differences in the Unmatched Brand-Exposure Rate

As shown in the figure, the left panel reports unmatched brand-exposure rates by platform interface, and the right panel reports rates by query category. In the left panel, n in parentheses is the number of brand-exposure instances for that platform interface; the right panel displays the 10 query categories with the highest rates.

## 3.4 Contact-Information Exposure and Differences by Query Formulation and Category

In addition to brands, this study analyzes business contact information, including telephone numbers, official websites, public social-media accounts, and addresses. The sample comprises 160,860 citations, 574 queries, and eight platform interfaces, with June 23, 2026 as the reference date. “Detected in source” means that the body of the cited source contains contact information, while “exposed” means that the answer contains contact information. The overall contact-information exposure rate was approximately 8.4% (95% CI [7.6%, 9.3%]); among cited sources containing contact information, 12.4% of the contact details appeared in answers.

**Table 3-3. Factor Importance for Contact-Information Exposure and Odds Ratios by Query Formulation**

| Factor | Importance | High-Exposure Query Formulation | Odds Ratio Relative to the Reference Group |
|---|---|---|---|
| Query Subtopic (`subcat`) | 0.56 | Brand-Related | Approximately 89× |
| Query Layer (`layer`) | 0.145 | Local-Life Purchase | Approximately 12× |
| Citation Role (`quote_role`) | 0.11 | Source-Request | Approximately 9.5× |
| Cited-Source Domain / Platform | 0.08–0.09 | (Semantic similarity) | Exposure OR = 6.2 |
| Recency / Character Count / Semantic Score / Sequence Number | 0.01–0.06 | — | — |

First, in the random forest's global permutation importance, query subtopic ranked first at 0.56. Brand-related queries had approximately 89 times the odds of contact-information exposure relative to the reference group ($p < 0.001$), followed by local-life purchase and source-request queries. Overall exposure rates were 8.32% for Web interfaces and 8.55% for App interfaces. In the multivariable logistic regression, the semantic similarity score had an odds ratio of approximately 6.2 ($p < 0.01$), making it one of the stronger continuous correlates. Permutation importance measures global predictive contribution at the model level, whereas odds ratios measure the strength of conditional associations after other variables are controlled; the two are therefore reported separately.

**Table 3-4. Comparison of Contact-Information Categories Detected in Sources and Exposed in Answers**

| Contact-Information Type | Detection Rate in Source Text (%) | Exposure Rate in Generated Answers (%) | Relative Exposure Ratio | Exposure Pattern |
|---|---|---|---|---|
| Social-Media Account (WeChat Official Account) | 7.1 | 29.8 | 4.20 | Substantially Overrepresented |
| Official Service Hotline | 6.4 | 23.2 | 3.62 | Substantially Overrepresented |
| Official Website or URL | 14.2 | 45.0 | 3.17 | Substantially Overrepresented |
| 400/800 Service Telephone Number | 15.3 | 19.7 | 1.29 | Slightly Overrepresented |
| Fixed-Line Telephone Number | 18.7 | 14.5 | 0.78 | Slightly Underrepresented |
| Mobile Telephone Number | 16.1 | 6.4 | 0.40 | Clearly Underrepresented |
| Email Address | 9.4 | 3.5 | 0.37 | Clearly Underrepresented |
| Detailed Address | 19.9 | 0.9 | 0.05 | Severely Underrepresented |

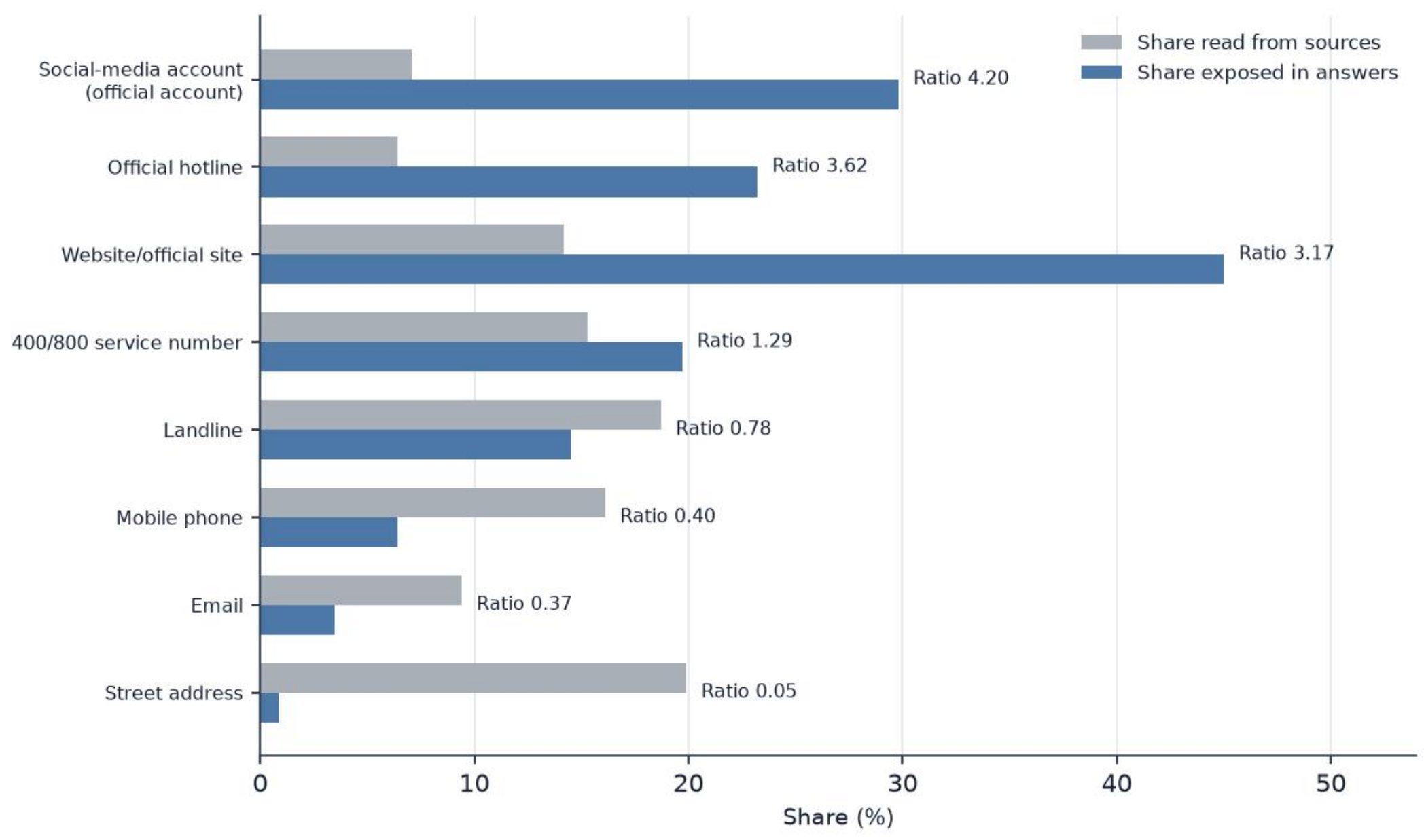

Figure 3-3. Shares of Contact-Information Categories Detected in Sources and Exposed in Answers

As shown in the figure, gray bars represent the share of each category detected in cited sources, and blue bars represent its share among contact information exposed in answers. The ratio following each bar is the answer-exposure share divided by the source-detection share.

### 3.4.1 Body-Text Crawl Coverage and Source-Matching Results

Table 3-4 shows that WeChat Official Accounts, hotlines, and official websites accounted for larger shares in answers than in the crawled body text, whereas mobile numbers, email addresses, and store addresses accounted for relatively smaller shares. Of the contact information exposed in answers, 28.9% could be matched in the crawled body text and 71.1% could not; the latter were uniformly labeled “unmatched.”

Under the current body-text extraction definition, the traceability rates for 400 service numbers and fixed-line telephone numbers were 61.7% and 55.1%, respectively, while those for URLs, WeChat Official Accounts, and hotlines ranged from 25% to 30%. URLs, public accounts, and hotlines may also occur in structured regions such as buttons, footer contact cards, or JSON-LD. Table 3-4 therefore reports the source-detection-to-answer-exposure ratio within the scope of body-text extraction. Specifically:

1. Under the current body-text crawl coverage, the answer-exposure-to-source-detection ratios for store addresses, mobile telephone numbers, and email addresses were 0.05, 0.40, and 0.37, respectively, indicating relatively low representation of these categories in answers. Under the current extraction definition, conversion from source detection to answer exposure was higher for official-channel contact information and lower for private-information categories such as mobile numbers, email addresses, and addresses.

The platforms also displayed clear source fingerprints. TYQW favored search-aggregated user-generated content; TYQWA favored ranked product guides; Tencent Yuanbao favored social media and online travel agencies; DeepSeek relied most heavily on official sources, particularly government and corporate official sites; and Doubao relied primarily on corporate official sites. These patterns are consistent with the platform heterogeneity reported in Section 2.1.

## 3.5 Platform-Native Commercial Components: Shopping Cards

Shopping cards are triggered by native interface components on particular platforms, and their distribution reflects differences at the platform-product and interaction levels. This section therefore examines the overall shopping-card trigger rate, platform distribution, and differences by query type.

In addition to textual exposure, some platforms display shopping cards in answers, including product cards, local-life cards, Taobao cards, and Damai cards. Among 32,702 valid rows, the overall trigger rate was approximately 4.8%. Cards were triggered on only two platform interfaces—DOUBA at 16.6% and TYQWA at 21.8%—and the trigger rate was 0 on the other six interfaces in this dataset. Triggering was strongly associated with shopping intent: comparison queries had the highest rates on both platforms, reaching 84.2% on DOUBA and 71.9% on TYQWA, followed by purchase/e-commerce queries.

**Table 3-5. Overview of Shopping-Card Triggers**

| Card Type | Platform | Sample Size (n) | Principal Characteristics |
| --- | --- | --- | --- |
| Product Card | DOUBA | 263 | Primarily multi-product comparisons; a relatively large mean number of cited sources (16.4) and comparatively concise answers. |
| Local-Life Card | DOUBA | 40 | Focuses on localized, in-depth guides to lifestyle services; longest mean answer length (2,866 Chinese characters). |

| Card Type | Platform | Sample Size (n) | Principal Characteristics |
|---|---|---|---|
| Taobao Card | TYQWA | 395 | Primarily e-commerce decision support and product recommendations; few independent cited sources, with a mean of approximately 0.4 citations. |
| Damai Card | TYQWA | 3 | All three Damai-card triggers came from the "industry dimension—leisure and entertainment" query group; the trigger rate in this group was 5.0% (3/60). |

The shopping-card analysis reveals four features. First, the component was observed on only two platform interfaces, indicating clear platform specificity. Second, Taobao cards frequently corresponded to in-platform product links, with `total_cite` of approximately 0.4; this scenario primarily reflects in-platform product links rather than attributes of general web sources. Third, after video cards were classified as no card, the TYQWA trigger rate fell from approximately 82% to 21.8%, showing that incidence was highly sensitive to classification rules. Fourth, answer-level variables overlapped informationally with the target definition, producing a LightGBM AUC of 1.0 and indicating target leakage. This section therefore treats platform distribution, triggering scenarios, and differences in classification definitions as its principal analytical results.

## 3.6 Summary: Exposure as Secondary Selection and the Need to Distinguish Source Populations

Both brands and contact information narrowed markedly between cited sources and answers. The selection rate for citation-pool brands was 8.3%, and 12.4% of contact information from cited sources containing such information ultimately appeared in answers. The two proportions use different units of analysis and cannot be multiplied or compared directly, but both demonstrate that "appearing in a source" is not equivalent to "being exposed in an answer."

Brand and contact-information exposure showed systematic differences with cross-source occurrence count, degree of citation absorption, query type, and semantic similarity score. The 5118-Baidu Composite Quality Score was a moderately positive predictor in the brand-selection model but made a relatively small predictive contribution in the contact-information exposure model.

Unmatched exposures occurred for both brands and contact information, indicating that entities presented in answers did not correspond completely to the citation text visible at that time.

Brand and contact-information exposure were statistically associated with cross-source occurrence count, degree of citation absorption, semantic role, position, source type, and related variables, and some associations differed across platforms and interfaces. These results provide testable hypotheses for subsequent controlled testing.

Figure 3-4 summarizes the chapter's results on citation absorption and entity exposure.

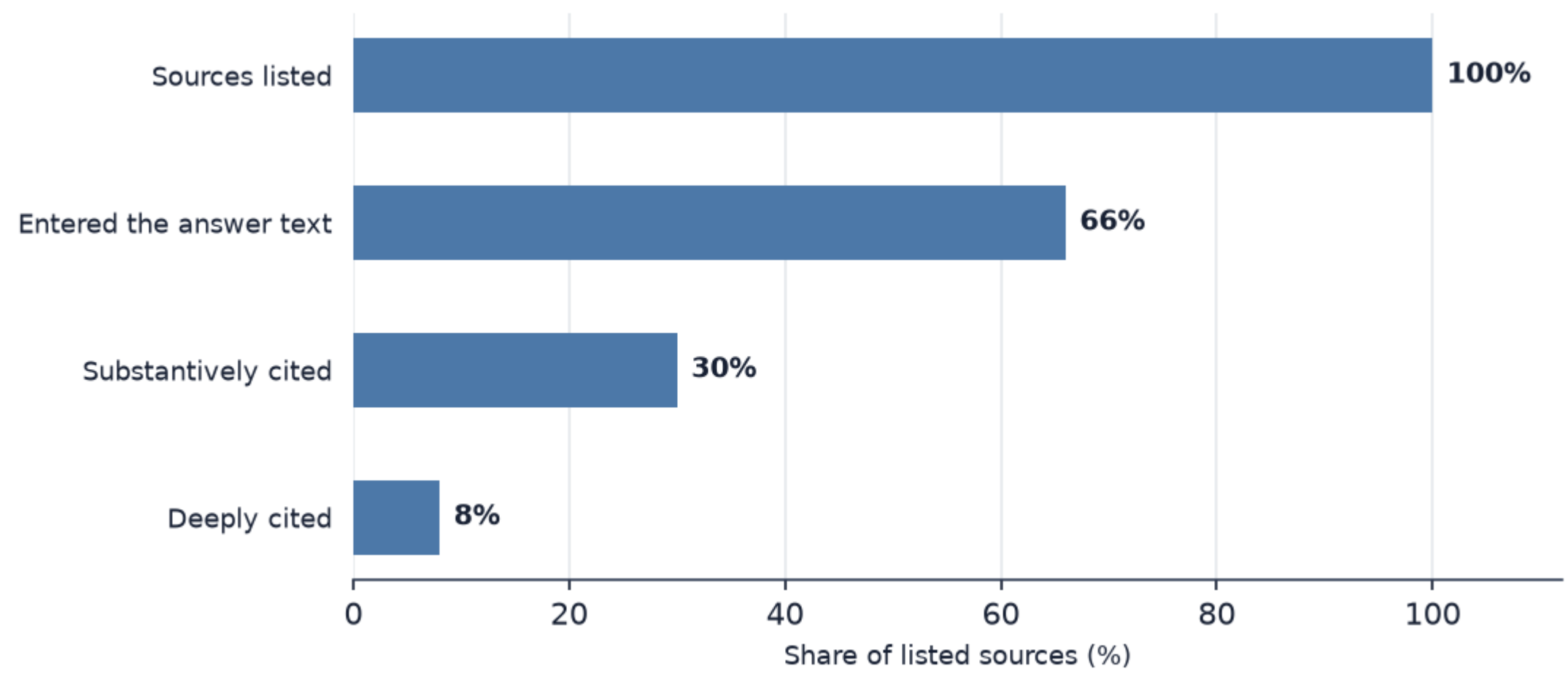

**Figure 3-4A. Citation-Absorption Proportions**

As shown in Figure 3-4A, all listed sources form the common denominator. Of all listed sources, 66% entered the answer body, 30% were classified as substantively cited, and 8% reached the level of deep citation.

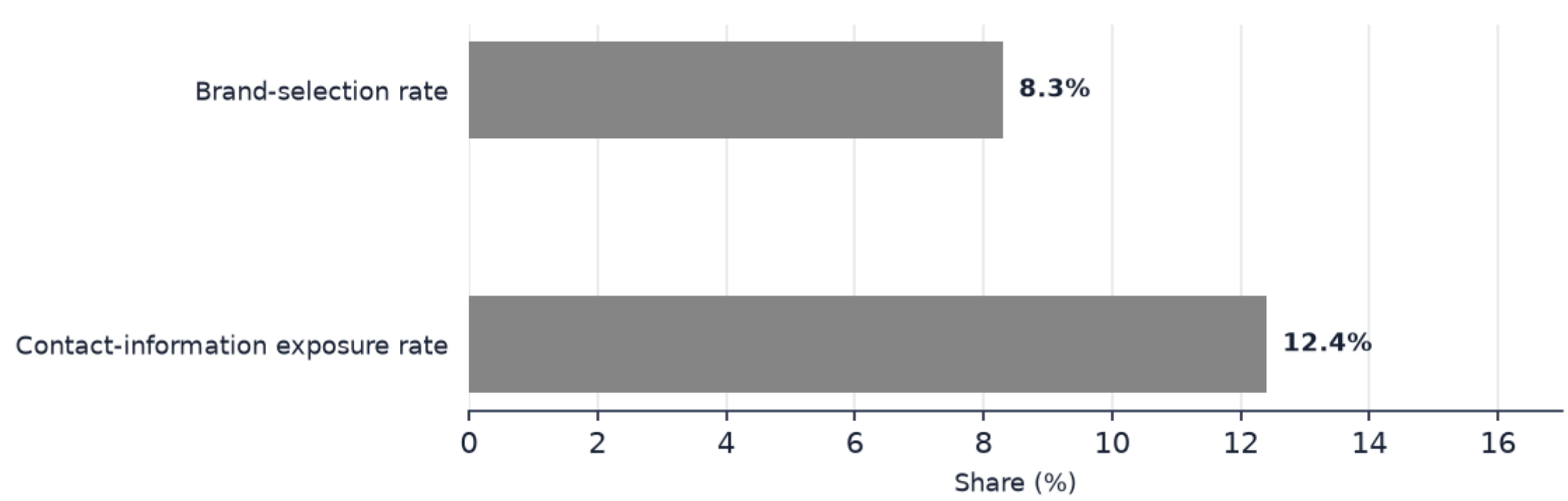

**Figure 3-4B. Brand- and Contact-Information Exposure Proportions**

As shown in Figure 3-4B, brand and contact-information exposure in answers are reported separately. The answer-selection rate for citation-pool brands was 8.3%; among cited sources containing contact information, the contact-information exposure rate was 12.4%. The denominators for the two indicators are citation-pool brands and cited sources containing contact information, respectively.

Chapters 2 and 3 report results on cited sources and entity exposure in answers, respectively. Chapter 4 further compares source sets, answer length, and citation counts between the App and Web interfaces of the same platform to test whether these results generalize directly across access interfaces.

# Chapter 4 Cross-Interface Consistency

This chapter pairs the App and Web interfaces of the same platform to compare the consistency of source sets, answer length, and citation counts and thereby examines how platform access interface stratifies the preceding results.

The App and Web interfaces of the same platform showed systematic differences in their cited-source sets, and between-platform differences were larger than differences among the six industry groups listed in Table 4-3. Cross-interface analyses must therefore be reported and validated separately by platform access interface.

## 4.1 Interface-Level Pairing Method

The analysis is based on the citation-level master table, comprising 160,860 rows and 614 queries. Platform interfaces are identified directly from `platform_code`: App interfaces are coded DOUBA, DPA, TYQWA, and TXYBA, and the corresponding Web interfaces are DB, DP, TYQW, and TXYB. The pairings are DB ↔ DOUBA, DP ↔ DPA, TYQW ↔ TYQWA, and TXYB ↔ TXYBA. Two methodological treatments are applied:

1. Interface-level union pairing. `run` is not treated as a separate dimension. For each query–interface combination, cited sources are first unioned across runs and then paired between the two interfaces, preventing random variation in a single collection from contaminating cross-interface differences.

2. Three-granularity consistency measurement. For each query, the Jaccard similarity of the source sets returned by the two interfaces—intersection divided by union (Vijaymeena & Kavitha, 2016)—is calculated at three granularities: exact URL (`quote_url`), registered domain (`domain`), and site name (`site_name`). Answer-level brand-aggregation fields are not included in this chapter.

A Jaccard coefficient of 1 indicates identical source sets across interfaces, whereas 0 indicates no overlap. The paired sample sizes are 580 for Doubao, 580 for DeepSeek, 591 for Tencent Yuanbao, and 352 for Qwen. Collection on the Qwen App interface was relatively sparse, resulting in fewer pairs; this asymmetry is itself an expression of cross-interface difference.

## 4.2 Source Consistency at Three Granularities: Only Approximately Half Overlap Even at the Domain Level

**Table 4-1. App–Web Source Consistency within Platforms at Three Granularities (Mean Jaccard Coefficients)**

| Platform | URL Level | Domain Level | Site-Name Level | Interpretation |
|---|---|---|---|---|
| DeepSeek | 0.408 | 0.507 | 0.504 | Relatively Most Consistent |
| Doubao | 0.310 | 0.505 | 0.491 | Relatively Most Consistent |
| Tencent Yuanbao | 0.097 | 0.299 | 0.214 | Moderately Inconsistent |
| Qwen | 0.014 | 0.190 | 0.120 | Almost No Overlap |

None of the four platforms had identical source sets across its App and Web interfaces. Mean domain-level Jaccard coefficients ranged from approximately 0.19 to 0.51, and URL-level consistency was lower still. The 95% confidence intervals for the mean domain-level coefficients were [0.490, 0.520] for Doubao, [0.491, 0.524] for DeepSeek, [0.289, 0.309] for Tencent Yuanbao, and [0.179, 0.200] for Qwen. Domain- and URL-level indicators jointly confirm that source sets only partially overlapped across interfaces.

For Qwen, the mean URL-level Jaccard coefficient was 0.014, 64.2% of queries had no exact URL in common across interfaces, and the mean domain-level coefficient was 0.19. For Tencent Yuanbao, the mean domain- and URL-level coefficients were 0.30 and 0.10, respectively. Cross-interface overlap was lower for Qwen and Tencent Yuanbao than for Doubao and DeepSeek, demonstrating clear platform heterogeneity in the magnitude of access-interface differences.

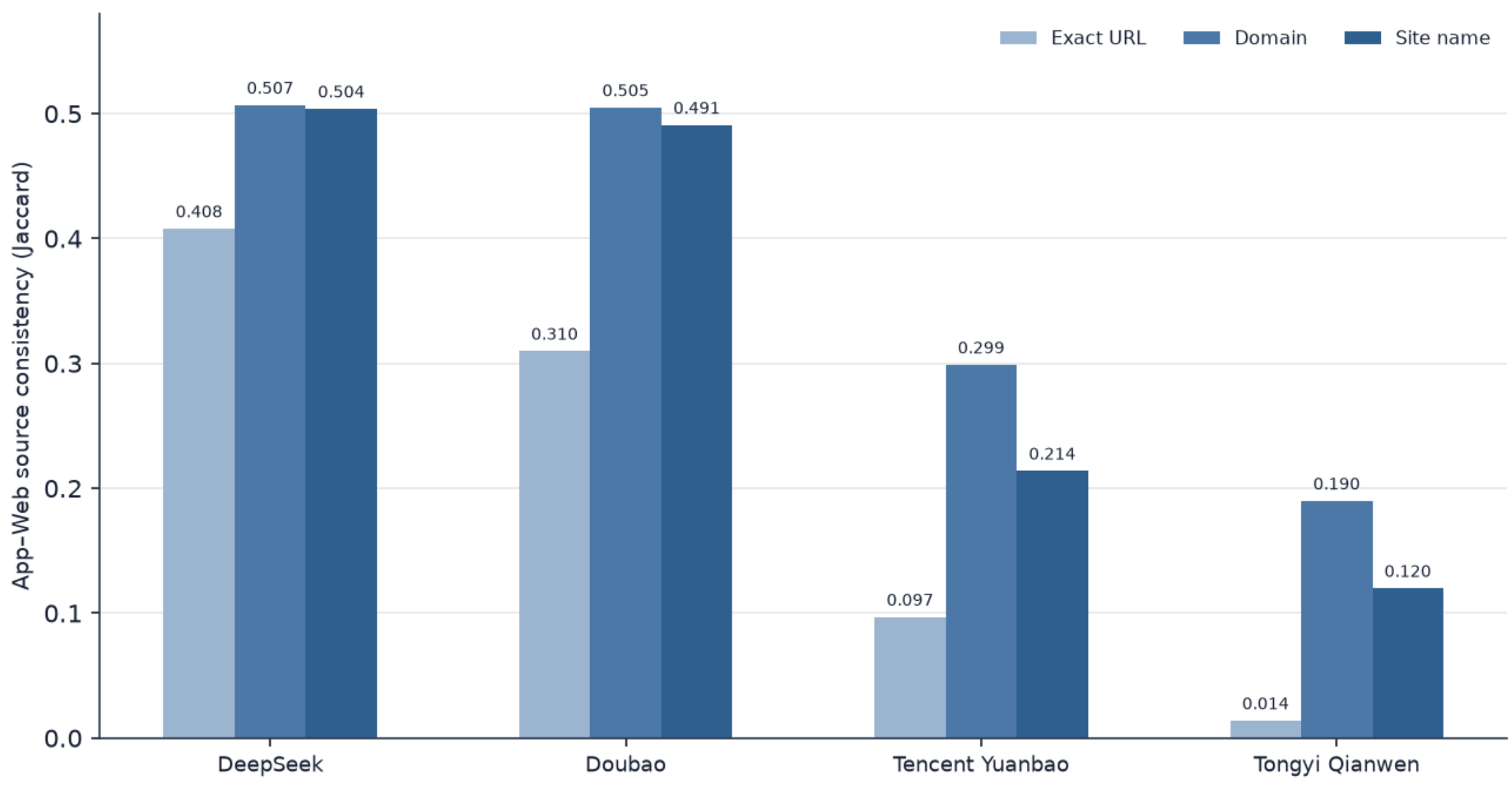

Figure 4-1. App–Web Source Consistency within Platforms at Three Granularities (Jaccard)

As shown in the figure, exact URL, domain, and site name represent source sets at the exact-URL, registered-domain, and consolidated-site-name levels, respectively. Bar height is the mean App–Web Jaccard coefficient, and error bars at the domain level show 95% confidence intervals.

## 4.3 Top-Ranked Sources and Answer Form: Differences Extend beyond the Long Tail

To test whether cross-interface differences arise primarily from long-tail sources, this study further compares the overlap of the top three sources on each interface together with answer length and citation count. Results are shown in Table 4-2.

**Table 4-2. Cross-Interface Differences in Top-Source Overlap and Answer Form**

| Platform | Top-3 Source Overlap | Web Answer Length | App Answer Length | Web Citation Count | App Citation Count |
|---|---|---|---|---|---|
| DeepSeek | 0.484 | 1,792 | 1,157 | 9.2 | 9.2 |
| Doubao | 0.457 | 1,416 | 1,394 | 18.5 | 18.1 |
| Tencent Yuanbao | 0.262 | 2,235 | 1,441 | 13.9 | 13.3 |
| Qwen | 0.133 | 2,074 | 1,286 | 20.1 | 11.0 |

DeepSeek had the highest top-three source overlap, at 0.48, whereas Qwen had an overlap of 0.13, indicating that differences also occurred among high-frequency sources. Answer length and citation count likewise differed across interfaces. For example, the mean answer lengths for DeepSeek Web and App were 1,792 and 1,157 Chinese characters, respectively, while Qwen Web and App returned 20.1 and 11.0 citations per answer, respectively. These values show that access-interface differences simultaneously affect high-frequency sources, answer length, and citation count.

## 4.4 Between-Platform Differences Exceed Differences among Six Industry Groups

To compare dispersion across the platform and industry dimensions, Table 4-3 reports six industry groups with paired samples, whose cross-interface Jaccard coefficients ranged from 0.375 to 0.434.

**Table 4-3. Cross-Interface Consistency by Industry (Mean Domain-Level Jaccard Coefficients)**

| Industry Category | Sample Size (n) | Mean Metric Value |
|---|---|---|
| Beauty and Hairdressing | 66 | 0.434 |
| Leisure and Entertainment | 72 | 0.432 |
| Wedding Services and Photography | 72 | 0.427 |
| Food and Dining | 73 | 0.414 |
| Hotels and Homestays | 68 | 0.410 |
| Health and Wellness | 72 | 0.375 |

Across the six industry groups compared, the range of platform-level Jaccard coefficients was 0.32, exceeding the between-industry range of 0.059 and indicating greater discrimination along the platform dimension.

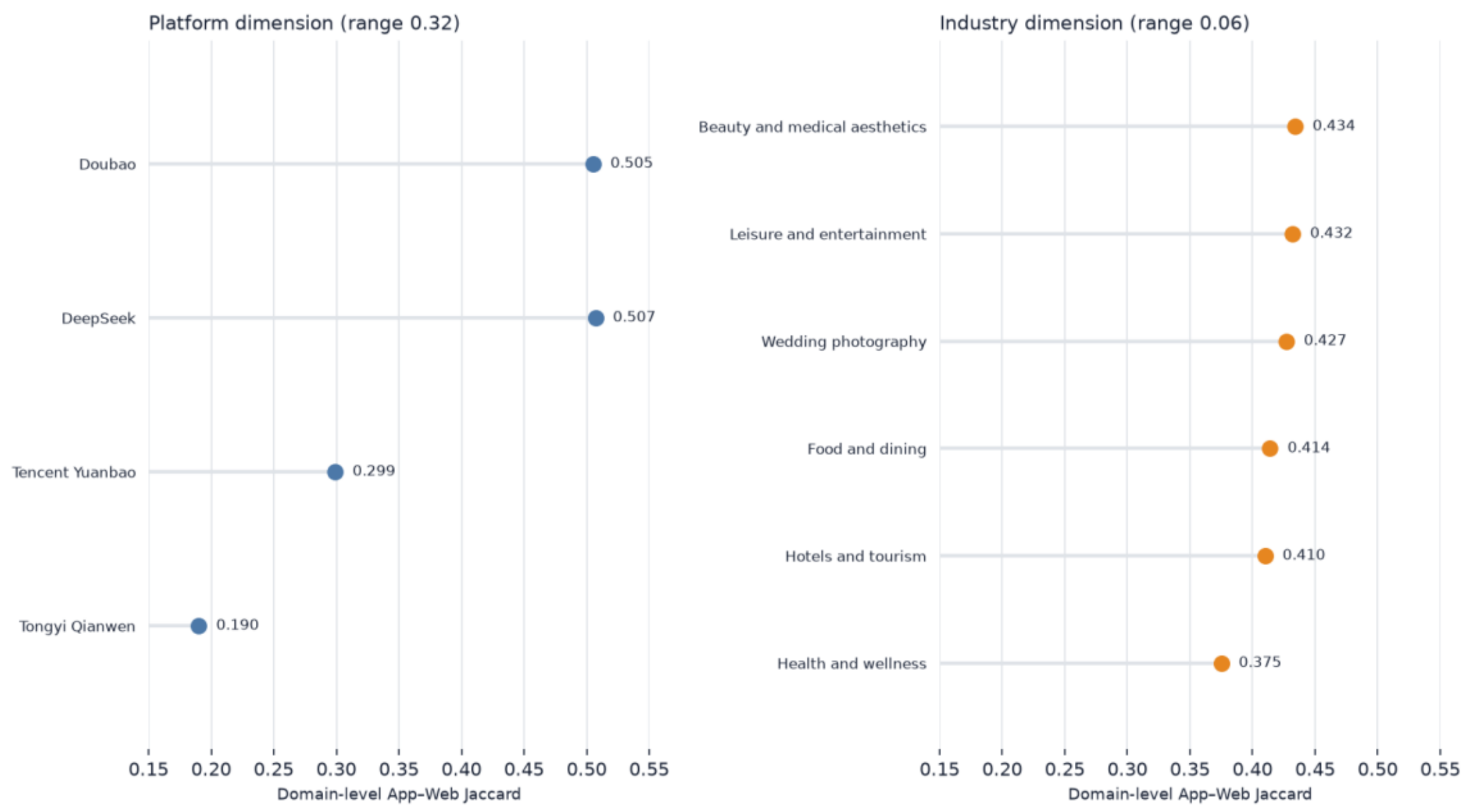

Figure 4-2. Distributions of Cross-Interface Consistency across Platform and Industry Dimensions

Note: The left panel reports mean domain-level App–Web Jaccard coefficients for Doubao, DeepSeek, Tencent Yuanbao, and Qwen. The right panel reports means for six industry groups: beauty and hairdressing, leisure and entertainment, wedding services and photography, food and dining, hotels and homestays, and health and wellness. The ranges across the platform and industry dimensions are 0.32 and 0.06, respectively.

## 4.5 Robustness Test: Cross-Interface Inconsistency Is Not an Artifact of Set Size

Jaccard similarity decreases mechanically when citation-set sizes differ across interfaces. To exclude this measurement artifact, two additional metrics that are not mechanically penalized by unequal set sizes are calculated. The first is the overlap coefficient—the intersection divided by the smaller set—also known as the Szymkiewicz–Simpson coefficient (Vijaymeena & Kavitha, 2016). The second is equal-length top-k Jaccard: the k most frequent domains are retained from each interface before Jaccard similarity is calculated, where k = min(|Web|, |App|). Results are shown in Table 4-4.

**Table 4-4. Set-Size Robustness Tests of Cross-Interface Source Consistency (Domain Level)**

| Platform | Jaccard | Overlap Coefficient | Equal-Length Top-k Jaccard | Unique Domains, Web / App |
|---|---|---|---|---|
| Doubao | 0.505 | 0.794 | 0.546 | 17.4 / 14.2 |
| DeepSeek | 0.507 | 0.730 | 0.505 | 14.7 / 13.4 |
| Tencent Yuanbao | 0.299 | 0.528 | 0.306 | 16.3 / 17.9 |
| Qwen | 0.190 | 0.429 | 0.190 | 12.0 / 7.4 |

After set size was controlled, equal-length top-k Jaccard coefficients remained close to the original coefficients: 0.19 for Qwen, 0.31 for Tencent Yuanbao, 0.55 for Doubao, and 0.51 for DeepSeek. Low cross-interface overlap therefore persisted for equal-length sets. The overlap coefficients for Doubao and DeepSeek ranged from 0.73 to 0.79, suggesting that much of their difference reflected asymmetric coverage of one interface by the other. The overlap coefficients for Qwen and Tencent Yuanbao were 0.43 and 0.53, respectively, indicating low overlap in both directions.

## 4.6 Summary: Systematic Cross-Interface Differences in Sources and Answer Form

Source sets differed systematically between the App and Web interfaces of the same platform; cross-interface results must therefore be reported separately by platform access interface.

Cross-interface differences occurred not only in the full source sets but also in the top three sources, answer length, citation count, and related indicators, showing that access-interface heterogeneity spans both source selection and answer form.

Among the six industry groups listed, the magnitude of between-platform differences exceeded that of between-industry differences.

Combining indicators such as deep-citation rate, source concentration, cross-interface consistency, and unmatched brand-exposure rate, Figure 4-3 shows that the four platforms exhibit different configurations of characteristics, further illustrating multidimensional heterogeneity across platforms and interfaces.

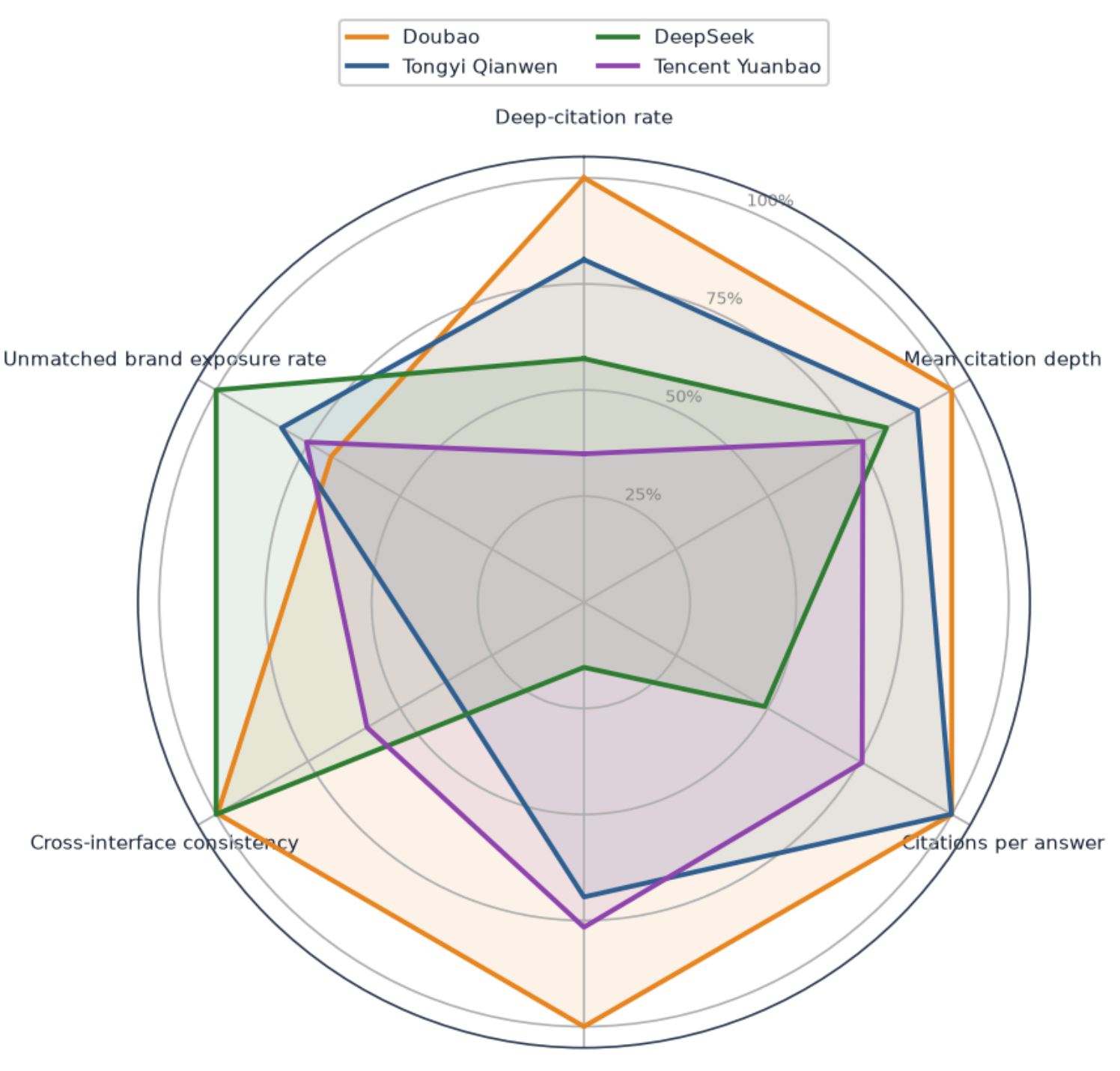

**Figure 4-3. Multidimensional Profiles of Citation Behavior across Four Platforms (Each Dimension Normalized to Its Maximum = 100%)**

Note: The six dimensions are deep-citation rate, mean citation depth, mean citations per answer, citation concentration (HHI), App–Web cross-interface consistency, and unmatched brand-exposure rate. Each dimension is normalized to the maximum among the four platforms as 100% to compare platforms' relative positions on that dimension. Doubao is relatively high in deep-citation rate, mean citation depth, mean citations per answer, and citation concentration, and its cross-interface consistency is close to that of DeepSeek. DeepSeek has the lowest citation concentration and highest cross-interface consistency, together with the highest unmatched brand-exposure rate. Qwen has a mean citation count close to Doubao's but the lowest cross-interface consistency. Tencent Yuanbao has relatively low deep-citation rate and mean citation depth, with cross-interface consistency between those of DeepSeek and Doubao and that of Qwen. The figure label "brand without-source rate" corresponds to the consistently defined "unmatched brand-exposure rate" used in this study.

Chapter 4 incorporates platform and access-interface heterogeneity into the stratified interpretive framework for the results in Chapters 2 and 3. The next section integrates the findings on citation, exposure, and consistency.

# Discussion: Principal Findings and Research Implications

Synthesizing Chapters 2–4, this study treats cited sources, degree of citation absorption, entity exposure, and cross-interface consistency as interrelated analytical levels with different units of analysis. The sample results show that content fit, cross-source occurrence count, article recency, degree of citation absorption, and platform access interface are associated with outcome variables at different stages, whereas the 5118-Baidu Composite Quality Score was not the leading predictor in most models. The following discussion develops testable applied hypotheses around these association structures.

## Principal Findings and Their Research Implications

The research implications are discussed from four perspectives: source ecology, article recency, query formulation and contact-information exposure, and platform access interface.

Cross-source occurrence count showed a stable gradient with brand-selection rate and should therefore be prioritized for testing in subsequent controlled experiments.

The brand-selection rate associated with deep citations was 13.7 times that associated with nominal citations, while semantic roles such as comparisons, ranked lists, definitions, and data had selection rates above the overall baseline. Differences among semantic roles support incorporating content structure and evidence organization into subsequent optimization experiments to test their effects on citation absorption and brand exposure.

The article-recency results can be translated into a temporal-scale hypothesis for future longitudinal research. For high-timeliness queries, the fitted half-life of the relative frequency of cited pages was approximately 39 days; that is, relative frequency was approximately halved for each additional 39 days of publication-age distance. This half-life provides a testable temporal benchmark for subsequent longitudinal studies.

Contact-information exposure was statistically associated with query category, semantic similarity score, and source type. Brand-related queries had an odds ratio of approximately 89 relative to the reference group, and the semantic similarity score had an odds ratio of approximately 6.2. Significant differences across brand-related queries and source types indicate that task intent is the primary stratification factor for contact-information exposure. Future research can incorporate these variables into preregistered controlled designs to test their replicability and causal direction.

Source concentration, semantic similarity scores, and article-recency distributions all differed across platforms and access interfaces; the relevant indicators should therefore be reported separately for each specific platform interface.

The unmatched brand-exposure rate varied systematically with total brand frequency, citation-pool size, and query type, indicating identifiable structural differences and providing priority cases for subsequent provenance analysis.

## Distinguishing the 5118-Baidu Composite Quality Score from Institutional Source Type

In analyses of degree of citation absorption, semantic role, within-answer position, and publication-age distance, the predictive importance or correlation of the 5118-Baidu Composite Quality Score ranked relatively low. In the brand-selection model, it was a moderately positive predictor, indicating that its predictive contribution varied by analytical task. Institutional source composition also differed between YMYL and consumer domains. The 5118-Baidu Composite Quality Score and institutional source type are distinct constructs and should be used separately to measure traditional SEO performance and source-institution attributes.

## Measurement Design and Future Research

App–Web source-set differences and missingness patterns in the publication-time field indicate that future research must incorporate platform, access interface, and metadata availability into measurement design. Future studies can supplement the data with internal platform retrieval records, full-page structured content, and interactive elements to test the technical sources of interface differences and further classify the provenance of unmatched exposures.

# Research Limitations and Scope of Applicability

The principal limitations concern identification design, sample coverage, measurement error, and statistical analysis. The following qualifications define the scope within which the results apply.

Identification design. This study primarily analyzes observational data. Although queries were deployed according to prespecified `layer–subcat` combinations, no single-factor randomization was implemented. Relationships among source attributes, article recency, semantic role, platform characteristics, and outcome variables should therefore be interpreted as conditional associations within the query set and platform scope examined here.

Sample and generalizability. The study observes webpages that had already entered citation lists; it did not obtain the complete candidate pool or rejected pages from the open Web. Data collection was concentrated in June and July 2026 and covered four platforms, eight platform interfaces, and 614 controlled queries. The conclusions reflect this collection window and sample scope, while cross-interface industry comparisons are limited to the six industry groups with paired samples.

Measurement error. Citation-absorption type, semantic role, and semantic similarity score were generated by a large language model under fixed rules and are operational measurements. Extraction of cited body text excluded structured regions such as pop-up windows, buttons, footer contact cards, and JSON-LD, potentially underestimating the presence of URLs, WeChat Official Accounts, hotlines, and similar information on source pages. Contact-information matching rates and source-detection-to-answer-exposure ratios therefore reflect the body-text extraction definition. In addition, the publication-time field was systematically missing on some platform interfaces, and the article-recency analysis is based on the subsample with available publication times.

Statistical and model boundaries. The large sample increases the ability to detect small differences. Accordingly, the study relies primarily on effect magnitude, direction, and robustness and distinguishes exploratory tests from confirmatory inference. Some platform-native components appeared on only a few platform interfaces and are treated as supplementary descriptive results. Models exhibiting information leakage because their features overlap with the target definition are excluded from the core inferences.

# Conclusion

This study constructed a unified corpus from eight platform interfaces, 614 queries, and three fixed collection replications. The raw data comprised 214,119 records, and the cleaned citation-level master table comprised 160,860 records. The analyses of citations, entity exposure, and cross-interface consistency yield four principal conclusions:

First, among sources listed as citations, both absorption of source content into answers and inclusion of source-mentioned brands in answers were selective. The proportion of citation-pool brands entering answers was 8.3%. This overall selection rate indicates strong secondary selection between citation and brand exposure.

Second, content fit, cross-source occurrence count, and degree of citation absorption were principal predictors in the corresponding models. The performance of the 5118-Baidu Composite Quality Score varied by analytical task, and institutional source type and traditional SEO performance are distinct constructs.

Third, article recency was only weakly correlated with the 5118-Baidu Composite Quality Score, and the recency distribution of cited pages varied with query time sensitivity. The fitted half-lives for high- and low-timeliness queries were approximately 39 and 68 days, respectively, indicating faster decay and a more recent distribution of cited pages for highly time-sensitive queries.

Fourth, heterogeneity existed across platforms and access interfaces. Source sets were not identical between the App and Web interfaces of the same platform; among the six industry groups listed in Table 4-3, between-platform differences exceeded between-industry differences.

Overall, this study identifies systematic association structures linking cross-source occurrence, citation absorption, article recency, contact-information category, and platform access interface with entity exposure. Future research can test causal effects using randomized controlled probes, collect additional data on page-update times, full-page structured content, and interactive elements to verify the provenance of unmatched exposures, and assess temporal stability through repeated collection across time points. Subsequent studies can also combine layout, retrieval traces, and generation-process data to examine the mechanisms producing early-position citations and whether repeated mention across sources constitutes a credibility signal used by platforms.

# Data and Research-Materials Availability

The GitHub repository `WENDAOstudy/cn-geo-citation-dataset` releases only the controlled-collection raw data and field documentation that can be made public for this study. The full crawled text of cited webpages and the topic-specific analysis scripts are copyrighted materials retained by the research team; they are outside the public release and are not available through individual access requests. Signals used in the 5118-Baidu Composite Quality Score were obtained from the 5118 Marketing Big Data Platform (https://www.5118.com/) and the Baidu Webmaster Platform (https://ziyuan.baidu.com/site/index). The public data retain only the derived fields required for the analyses in this paper and do not include restricted API credentials. https://github.com/WENDAOstudy/cn-geo-citation-dataset